\documentclass[preprint]{aastex63}

\usepackage{amsmath}
\usepackage{amssymb}
\usepackage{verbatim}

\newcommand{\oblLOS}{\psi_\mathrm{LOS}}
\newcommand{\oblsky}{\psi_\mathrm{sky}}
\newcommand{\rotrate}{\omega_\mathrm{rot}}

\received{2021 December 14}
\revised{2022 September 27}
\accepted{2022 October 31}
\submitjournal{AAS Journals}

\shorttitle{Eclipse Mapping and Rotation}
\shortauthors{Adams \& Rauscher}

\begin{document}

\title{The Sensitivity of Eclipse Mapping to Planetary Rotation}

\correspondingauthor{Arthur D. Adams}
\email{arthura@ucr.edu}

\author[0000-0002-7139-3695]{Arthur D. Adams}
\affiliation{University of California, Riverside}
\affiliation{University of Michigan}

\author[0000-0003-3963-9672]{Emily Rauscher}
\affiliation{University of Michigan}


\begin{abstract}
Mapping exoplanets across phases and during secondary eclipse is a powerful technique for characterizing Hot Jupiters in emission. Since these planets are expected to rotate about axes normal to their orbital planes, with rotation periods synchronized with their orbital periods, mapping provides a direct correspondence between orbital phase and planetary longitude. We develop a framework to understand the information content of planets where their rotation states are not well constrained, by constructing bases of light curves across different rotation rates and obliquities that are orthogonal in integrated flux across secondary eclipse. These demonstrate that brightness variability during eclipse may arise from a variety of rotation rates, obliquities, and map structures, requiring priors to properly disentangle each of these components. By modeling eclipse observations of the Warm Jupiter HAT-P-18 b we demonstrate that, at a signal-to-noise equivalent to $\sim 10$ orbits with JWST, confusion about map structure is likely a concern only at the upper physical limits of possible rotation rates. Even without priors, one may nevertheless be able to put an order-of-magnitude constraint on rotation rate by determining at what rates the fitted map complexity is minimized, a prescription whose efficacy increases if out-of-eclipse data are available to isolate the effects of rotation. Finally, in the limit of maps with longitudinal symmetry, the projected obliquity in the plane of the sky determines the information available during eclipse, ranging from non-detections of structure to a basic constraint on hemispherical asymmetry and orientation depending on the obliquity angle.
\end{abstract}

\section{Introduction} \label{sec:intro}
Mapping exoplanets has been a major development in the characterization of their atmospheres. By exploiting the geometry of the way one observes an exoplanet as it rotates and orbits its host star, one can pull out multiple spatial dimensions of information about the planet's surface brightness from a 1-dimensional variation of brightness with time. Mapping therefore bridges the gap between the inference of fundamental bulk properties of exoplanets --- including mass, radius, and equilibrium temperature --- and the eventual goal of directly resolving their features.

Planetary mapping in its most general form encompasses any structure revealed through the planet's spin, orbit, or spectrum \citep[for a review, see][]{Cowan2018}. Orbital phase curves, for example, map exoplanets longitudinally as successive hemispheres are made visible through a planetary orbit \citep[for a review, see][]{par17}. This style of mapping reveals structures such as day-night brightness contrasts \citep[e.g.~][]{knu07b,Demory2016}. \citet{Williams2006} and \citet{Rauscher2007b} first evaluated the feasibility of using the occulting edge of the star as a natural mask of an exoplanet during secondary eclipse; \citet{dew12} and \citet{maj12} were the first to put the technique into practice to make 2-dimensional exoplanet emission maps.

Mapping also constrains properties of the atmosphere by comparing the spatial distribution of brightness with the predicted structure of outgoing radiation from atmospheric circulation models \citep[e.g.~][]{Menou2009,Showman2010,Showman2020,Leconte2013,heng15,amu16,Wolf2017,Tan2019}. Many of these results come from phase curve mapping of close-in giant planets (``Hot Jupiters''), with phase variations probing their thermal and reflective (albedo) spatial structures. One notable example is inferring longitudinal offsets of the hottest emitting regions of a planet through offsets in time of the observed minimum and maximum of its phase variations \citep[e.g.~][]{knu09b,knu12,Demory2013,zel14,zha18,Schlawin2018}. There have been several efforts to explore the reach of what one can infer via mapping for observations both present and future \citep[e.g.,][]{cow08,Luger2021}, as it provides a capability to map certain brightness structures that are unique to the technique. For example, for a planet that gets eclipsed with an orbit at any angle other than purely edge-on ($i \neq 90^\circ$), any differences in brightness by observed latitude will manifest as a change in the eclipse shape \citep{Rauscher2007b,dew12,maj12,Cowan2018}. This provides the first piece of information about the planet's brightness structure unlocked uniquely by mapping through occultation. Beyond this information, it is helpful to understand as much as possible about the limitations of the additional information available through eclipse mapping, to determine where independent constraints of brightness maps can most usefully complement our understanding of planetary maps.

Many of the exoplanets with current mapping constraints are Hot Jupiters, whose rotations are assumed to be synchronized with their orbits. This provides a one-to-one correspondence between their orbital phases and the longitudes we observe at any given time. Rotation rates of planetary-mass companions have to date only been measured for young, giant planets on wide orbits, using Doppler broadening in their spectra \citep{Snellen2014,Bryan2020b,Xuan2020}; this method has been shown to be far less constraining of Hot Jupiter rotation rates, due to some degeneracy with broadening from atmospheric winds \citep{Flowers2019,Beltz2020}. However, for orbital periods longer than those of Hot Jupiters, the expected synchronization time scales for planets due to stellar tides reaches the typical age of a mature system\footnote{The exact orbital distance where this occurs depends highly on the precise system parameters. For a Jupiter-mass planet on a 10-day orbit around a solar-type star, initially rotating 100 times per orbit, the tidal locking time scale is $\sim 10^8$ years \citep[see the calculations for tidal locking timescales in][]{Gladman1996}}.

A hard limit on rotation rates is set by the break-up velocity, faster than which the speed of material at the planet's measured radius reaches the escape velocity for the planet \citep[e.g.~][]{Porter1996}:
\begin{align}\label{eq:break-up_velocity}
    v_\mathrm{b} &\sim \sqrt{\frac{2GM_\mathrm{P}}{3R_\mathrm{P}}} \\
    P_\mathrm{b} &= \frac{2 \pi R_\mathrm{P}}{v_\mathrm{b}} \\
    &\sim \sqrt{\frac{9 \pi}{2G\rho_\mathrm{P}}} \\
    &\approx 3.5~\mathrm{hr} \left(\frac{\rho_\mathrm{P}}{\rho_\mathrm{J}}\right)^{-1/2},
\end{align}
where $M_\mathrm{P}$, $R_\mathrm{P}$, and $\rho_\mathrm{P}$ are the planetary mass, radius, and density, respectively. $\rho_\mathrm{J} \approx$1.3 g cm$^{-3}$ is the bulk density of Jupiter. \citet{Bryan2020b} analyzed rotational line broadening in the spectra of 27 planetary-mass objects and compared their derived rotational velocities in terms of the break-up velocity, with measured fractions $v/v_\mathrm{b}=$0.05--0.66 in their sample. We currently do not have reliable measurements of rotation periods specifically for planets at intermediate orbital periods --- those too close to their stars to be directly imaged but too far to be easily characterizable targets with high-resolution transmission spectroscopy. The diversity of rotation rates of planets in our own Solar System points to a potentially large range, but as with the argument for widely-separated companions there should be a hard limit where the planet material must remain bound.

Complementary to the question of rotation rate is the alignment of the spin axis relative to the orbital plane, often measured as obliquity, which along with the rotation rate is expected to asymptotically dampen to a minimum energy state. In the case of spin axis orientation, the preferred obliquity is zero for a similar range of orbital periods as for spin synchronization. Just as with rotation rate, obliquities of Solar System planets span the entire possible range, and are thought to arise from a combination of tidal interactions between planets \citep[see e.g.~][]{Ward2004,Hamilton2004,Brunini2006} or past impacts \citep{Lissauer1991,Dones1993,Brunini1995}. To date, obliquity constraints have only come from widely-separated planetary-mass companions \citep[e.g.~][]{Bryan2020a,Bryan2021}, and both show significant spin obliquities which are likely to have been imprinted through the companions' formation pathway, potentially through gravitational instability in their protoplanetary disks. This represents the cutting edge of obliquity measurements, and therefore there are still many measurements left to the future for planets at a wide range of orbital separations, especially those thought to form from core accretion. Obliquity in planets on close-in orbits can also manifest itself in thermal phase curves, as was speculated for the Hot Jupiter CoRoT-2 b in \citet{Adams2019b}.

With these uncertainties outlined, there remains a question of how effective mapping techniques will be in the regime where rotation vectors cannot be assumed and so our ability is inherently limited in translating observed brightness variations to a well-defined set of coordinates on a planet's surface. This paper aims to address these questions, starting in \S \ref{sec:methods} where we describe the method of constructing a basis of ``eigenmaps'' which produces a orthogonal (or nearly orthogonal) basis of light curves around eclipse. In \S \ref{sec:eigenbases} we then explore how the structure of the basis changes as the rotation rate increases from the nominal expectation of spin-orbit synchronization, and additionally as obliquity increases. \S \ref{sec:retrieval} details a mock ``retrieval'' of a simple hotspot map across a variety of scenarios where in each we assume some fixed rotation for our model, and show the efficacy of fitting eclipse curves from planets rotating across various rotation rates and obliquities. We summarize our findings and discuss the potential of this mapping framework for the next generation of eclipse observations in \S \ref{sec:conclusions}.

\section{Constructing Eigenbases}\label{sec:methods}
Eclipse light curves provide information about map sub-structure principally during ingress and egress, where partial occultations occur. This is a projection of a 2-dimensional quantity (the net outgoing radiation from the planet in latitude and longitude) down to a 1-dimensional time series of integrated observed brightness. As such, it introduces a degeneracy. If we were able to look at the outgoing radiation from all angles, spherical harmonics would be a good choice for classifying what we observed, as that represents an orthonormal basis for information on the surface of a sphere, where each mode (commonly referred to using $\ell$ and $m$ subscripts) are mutually orthogonal and are ordered from largest to smallest variance across the surface at fixed normalization. For our observations, we want to find the closest to an orthonormal basis as we can in our special 1-dimensional time series. This depends on the nature of the function that represents how the spherical brightness map gets projected to the eclipse curve. The \href{https://luger.dev/starry}{STARRY} code \citep{lug19} provides these functions in its occulation code and allows us to generate simulated light curves for planetary systems with arbitrary brightness maps. STARRY indeed uses spherical harmonics to encode its maps; however, the eclipse curves that result from each mode of spherical harmonic map are themselves not mutually orthogonal. Therefore, to recover as orthogonal a basis of eclipse curves as is possible, we must transform our original spherical harmonic basis of maps into a new basis which generates the desired maximally orthogonal curves.

To do this we start with an initial spherical harmonic basis, where each map is a combination of a uniform component (corresponding to our assumed luminosity given by $L_\mathrm{P}/L_\star$ in Table \ref{table:warmJupiter_parameters}), and a single, distinct spherical harmonic mode with a weight equal to the constant term\footnote{Since all spherical harmonics other than the constant $Y_{00}$ term integrate to zero on the surface of a sphere, this means that adding harmonics will not affect the total luminosity.}. We generate our eigenmaps from harmonics up to degree $\ell=20$, and for simplicity we assume our model planet has a Jupiter mass and radius, and orbits a star with a solar mass and radius. To get the corresponding input curve from each map, we subtract the brightness curve that would come from a planet with the uniform brightness alone. We use this set of ``basis curves'' in a principal component analysis \citep[PCA; for examples of applications specific to exoplanet science see][]{Davis2017,Rauscher2018,Damiano2019}. PCA is a technique which can be used to numerically approximate a (nearly) orthogonal basis of light curves, themselves generated from maps which are some linear combination of spherical harmonic maps. \citet{Rauscher2018} first described applying PCA to generate a set of basis light curves, where it was used to quantify the information content of planetary surface brightness encoded in eclipse curves for hot Jupiters. There they referred to the new basis of light curves as ``eigencurves'', each of which were the result of ``observing'' a corresponding planetary map, also known as an ``eigenmap''. These terms reflect the way in which PCA projects the input data onto a new basis of eigenvectors, sorted by their eigenvalues. We refer to the new basis as a whole as the planet's ``eigenbasis'', where each ``mode'' of the eigenbasis has a map-curve pair. The first mode of the eigenbasis yields an eclipse light curve (the first eigencurve) which contains the highest variance relative to the light curve of a planet of uniform brightness. The second mode is projected to be orthogonal to the first eigencurve in such a way that maximizes its own variance, the third orthogonal to the previous modes and maximizing its own variance, and so on down the modes. The dimension of the light curve ``space'' is set by the dimensionality of the maps, which is itself set by the number of degrees used in the original spherical harmonics. The outcome of this is a basis of eclipse curves tuned to a given system which maximizes the information content of the data in as few representative ``observations'' (i.e.~eigencurves) as is possible. 

The eigencurve approach involves separating out information which has physical priors from that which is left as ``agnostic'', which is left to be classified on structure alone. In the former category we have priors such as constraints on the orbital period, planet-to-star radius ratio, orbital inclination, eccentricity. The approach is well suited for Hot Jupiters, where we also have a constraint on the rotation period through the orbital period: there is very little ambiguity in the correspondence between time in the observation and the visible latitudes and longitudes. \citet{Rauscher2018} demonstrated that eclipse mapping can work for any type of orbit, as long as the orbital parameters are known to sufficient precision so as not to cause degeneracy with spatial structure in the planet map. However, we are interested in mapping the class of planets known as ``warm'' Jupiters --- giant planets with orbital periods $\sim 10$ days. This is the scale of orbital periods where arbitrary rotation rates and obliquities may exist in a mature system \citep[e.g.][]{Rauscher2017}. If rotation is under-constrained, the projection function now has additional unknowns, and the eigenbasis created assuming one rotation rate may be less optimal than one that would be generated at the (unknown) correct rotation rate. It is also possible that two rotation states produce similar eigencurves from different eigenmaps; that would mean that both have degeneracies, but there is an additional degeneracy introduced by rotation. By exploring how the eigencurves and eigenmaps change as a function of which rotation rate one assumes, we can begin to explore how limited our information is when rotation is not well constrained.

\begin{deluxetable}{lcc}[htb!]
\tabletypesize{\footnotesize}
\tablewidth{0pt}
\tablecaption{System properties for our ``prototypical Warm Jupiter'' model planet, used in \S \ref{sec:methods}--\ref{sec:eigenbases}, and the Warm Jupiter HAT-P-18~b \citep[largely derived from][]{Hartman2011}, whose mock observations are the topic of \S \ref{sec:retrieval}. $N_\mathrm{tot}$ is the number of rotations the planet makes across the total duration of secondary eclipse; $N_\mathrm{i,e}$ is the number of rotations the planet makes across the duration of a single eclipse ingress or egress.}
\tablehead{\colhead{Parameter} & \colhead{Warm Jupiter} & \colhead{HAT-P-18 b}}
\startdata
\cutinhead{Fixed}
$M_\star/M_\odot$ & $1$ & $0.77 \pm 0.03$ \\
$R_\star/R_\odot$\tablenotemark{*} & $1$ & $0.75 \pm 0.04$ \\
$M_\mathrm{P}/M_\mathrm{J}$ & $1$ & $0.197 \pm 0.013$ \\
$R_\mathrm{P}/R_\mathrm{J}$\tablenotemark{*} & $1$ & $0.995 \pm 0.052$ \\
$L_\star/L_\odot$ & $1$ & $0.27 \pm 0.04$ \\
$L_\mathrm{P}/L_\star$ & $10^{-3}$ & $\approx 2.3\times10^{-4}$ \\
$P_\mathrm{orb}$ (days) & 10 & $5.508023\pm6\times10^{-6}$ \\
$a$ (AU) & 0.09 & $0.0559 \pm 0.0007$ \\
$b$ & 0.66 & $0.324^{+0.055}_{-0.078}$ \\
$e$ & 0 & 0\tablenotemark{**} \\
$i$ ($^\circ$) & $85.7$ & $88.8 \pm 0.3$ \\
$\omega_\mathrm{rot}/\omega_\mathrm{orb} \rightarrow N_\mathrm{tot}=1$ & $\approx 71$ & $\approx 46$ \\
$\omega_\mathrm{rot}/\omega_\mathrm{orb} \rightarrow N_\mathrm{i,e}$=1 & $\approx 432$ & $\approx 345$ \\
$\omega_\mathrm{rot}/\omega_\mathrm{orb}$ break-up limit & $\approx 66$ & $\approx 17$ \\
\cutinhead{Variable}
$\omega_\mathrm{rot}/\omega_\mathrm{orb}$ & 1, 3, 10, 30, 100, 300 &  1, 3, 10, 30 \\
$\Delta \phi_\mathrm{tot}$ $\left(^\circ\right)$ & 5--1526 & 8--231 \\
$\Delta \phi_\mathrm{i,e}$ $\left(^\circ\right)$ & 0.8--250 & 1--30 \\
$\oblLOS$ ($^\circ$) & 0, 30, 60, 90 & \\
$\oblsky$ ($^\circ$) & 0, 30, 60, 90 & \\
\enddata
\tablenotetext{*}{The reference radii for the Sun and Jupiter are taken from Astropy \citep{ast13}: $R_\odot=6.957\times10^8$ m, $R_\mathrm{J}=7.1492\times10^7$ m.}
\tablenotetext{**}{HAT-P-18 b has a measured non-zero eccentricity, $e = 0.084 \pm 0.048$, but for our purposes we assume a circular orbit. The reported eclipse durations in Table \ref{table:HAT-P-18b_parameters} reflect the assumption of $e = 0$.}
\label{table:warmJupiter_parameters}
\end{deluxetable}

\subsection{The Synchronous Eigenbasis}\label{sec:methods:base_case}
\begin{figure*}
\begin{center}
\begin{tabular}{cc}
\includegraphics[width=7cm]{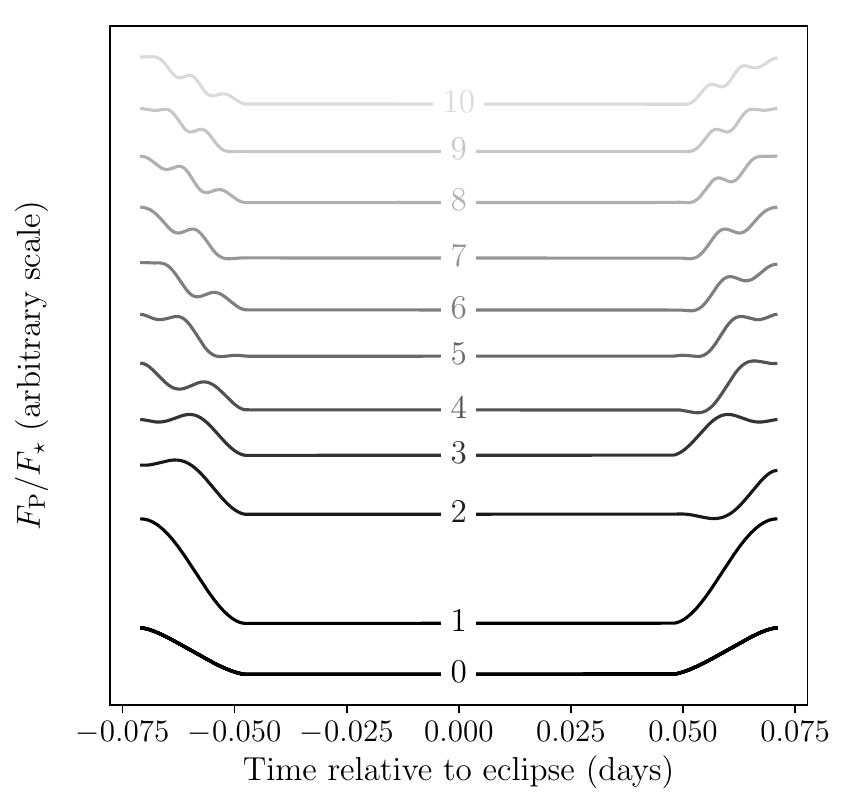} &
\raisebox{0.125\height}{\includegraphics[width=10cm]{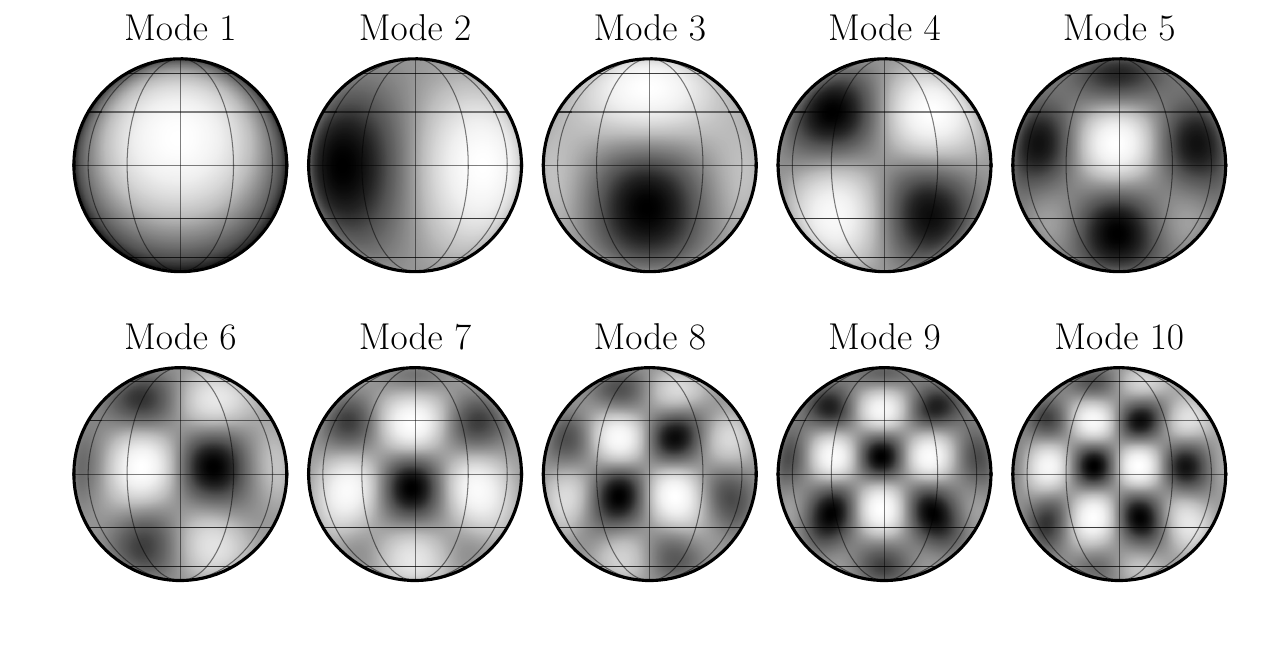}} \\
\end{tabular}
\caption{The first ten eigencurves identified by PCA for the zero-obliquity, spin-synchronous Warm Jupiter case. Brightness variations are approximately isolated to a single hemisphere, since the planet does not rotate very much over the duration of secondary eclipse. The maps represent the simplest 2-D patterns that produce sinusoid-like 1-D variations in the light curve during ingress and egress, with time ``wavelengths'' in half-multiples of the ingress and egress durations. Even and odd eigenmodes are symmetric and anti-symmetric, respectively, in reflection across the sub-observer longitude.}
\label{fig:base_eigenmodes}
\end{center}
\end{figure*}

As a first step, we show the synchronous eigenbasis, for a planet with a spin period equal to its orbital period (spin-orbit synchronization) and zero obliquity. This serves as the reference point for comparing how subsequent eigenbases differ as the rotation state is varied. Figure \ref{fig:base_eigenmodes} shows the eigencurves and eigenmaps constructed by our PCA routine. In both ingress and egress we obtain sinusoidal-like flux variations on top of the overall shape. The corresponding maps resemble the structures seen in spherical harmonics, albeit isolated to one hemisphere since the observable region during eclipse is approximately one half of the planet.

At this slow of rotation the eigenmaps are almost purely hemispherical --- the hemisphere which is swept across by the occulting edge of the star from our perspective. The shapes of ingress and egress for each mode resemble sinusoids, with an effective wavelength parametrized in terms of the ingress/egress duration. These wavelengths have a reasonably well-defined order, with the complexities of the eigenmaps increasing with each pair of modes. The even modes show more longitudinal structure, while the odds starting at mode 3 are more latitudinal. A further observation is that the modes appear to come in pairs, for example mode 2 with 3, 4 with 5, 6 with 7, etc. At eigenmode number $m$ we fit $\approx m/2$ wavelengths in a single ingress/egress, and the ingress is symmetric or anti-symmetric with egress depending on whether the mode is odd (latitudinal) or even (longitudinal), respectively. 

When faced with the task of estimating the 1-D light curve ``signal'' around eclipse, to try to recover information about the planetary brightness map, we can consider the idea of the Nyquist sampling limit. The amount of information for a sinusoid (or near-sinusoid) is set by the Nyquist limit, which requires an absolute minimum of $2n$ samples of a timeseries to constrain frequencies of at most $n$ per interval (in this case, ingress or egress duration). This implies that the absolute theoretical maximum number of eigenmodes one can hope to recover is approximately the number of samples per ingress or egress. We discuss this idea further in \S \ref{sec:conclusions}.

There are deviations away from this broad observation of symmetry or anti-symmetry between the ingress and egress of a given mode. Asymmetries can occur when there is neither symmetry nor anti-symmetry in the map structure observed in egress versus ingress. The primary symmetry here can be imagined as that of a reflection of the observed planet disk across a line which bisects the stellar disk perpendicular to the projected motion of the planet. Since the shape of the occulting edges for ingress and egress are mirrored relative to the disk of the star, if the observed portion of the map is different between ingress and egress, the shapes of the ingress and egress light curves will not be mirror images of each other. This effect will become increasingly important in subsequent cases.

\section{Eigenbases across Different Rotation States}\label{sec:eigenbases}
\begin{figure*}
\begin{center}
\begin{tabular}{cc}
\includegraphics[width=8cm]{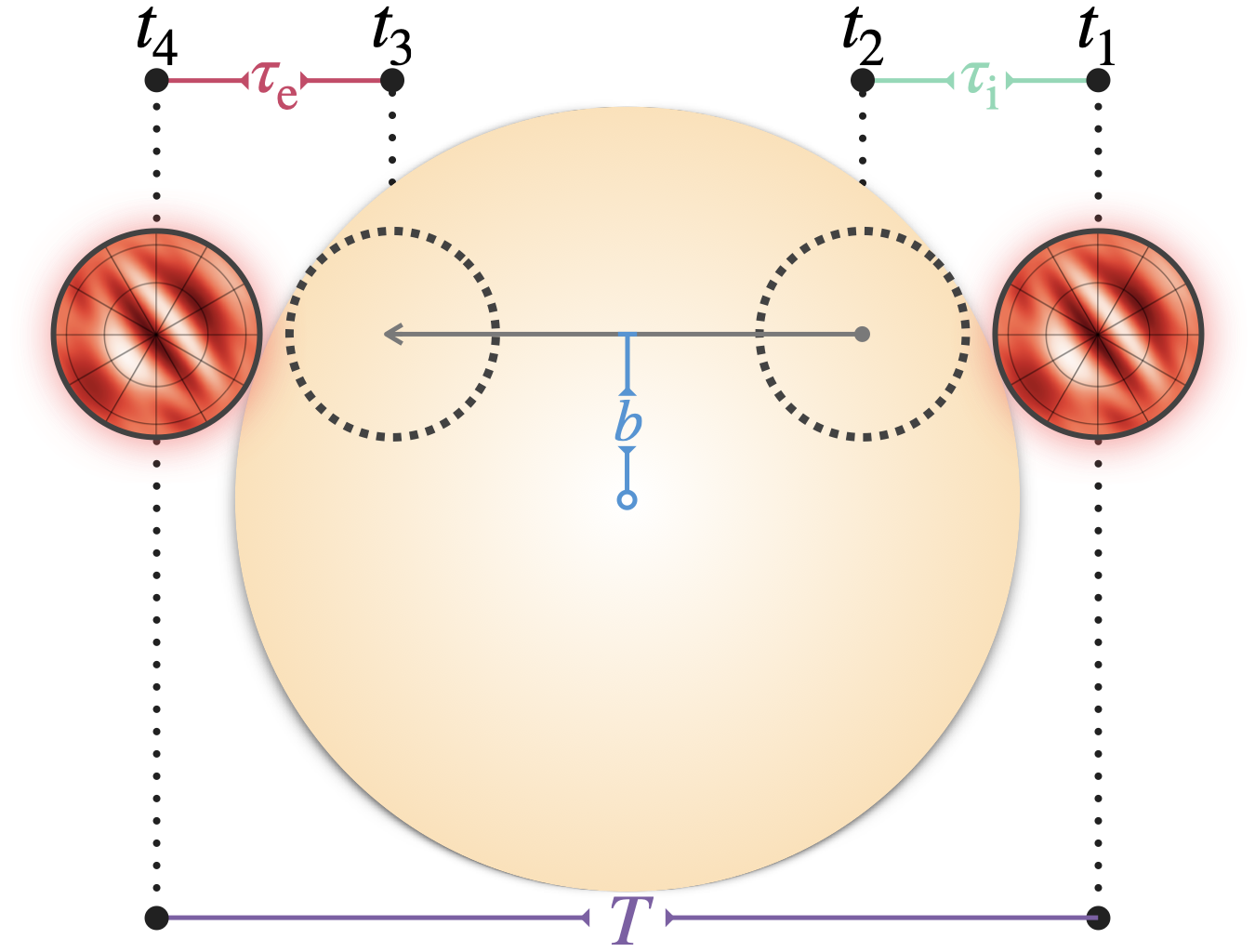} &
\includegraphics[width=8cm]{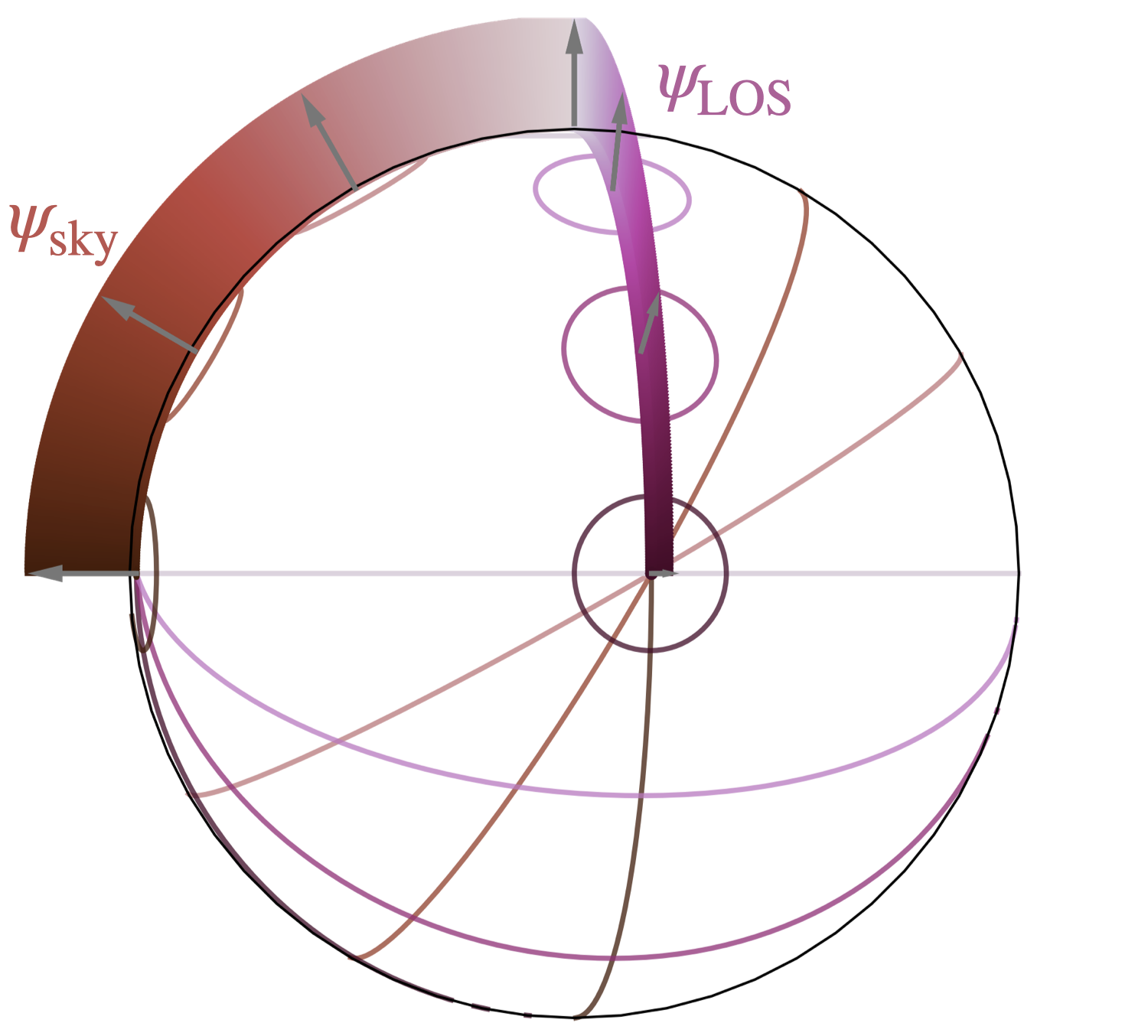} \\
\end{tabular}
\caption{Diagrams of the geometry and timing of eclipse, as well as the reference frame for the two dimensions defining the orientation of the planet's spin axis.}
\label{fig:system_diagrams}
\end{center}
\end{figure*}

For our warm Jupiter analysis we vary the rotation rate and the orientation of the spin axis relative to the line sight. This can be thought of as varying the components of the rotation vector
\begin{equation}\label{eq:rotation_vector}
    \vec{\omega} = \rotrate \left( \cos\oblsky \cos\oblLOS \hat{x} + \sin\oblsky \cos\oblLOS \hat{y} + \sin\oblLOS \hat{z}\right)
\end{equation}
where the plane of the sky is the $x$--$y$ plane and $\hat{z}$ points to the observer. $\hat{x}$ points to the ``right'', or from the stellar center to planetary center in a perfectly edge-on orbit ($i=90^\circ$) with zero longitude of the ascending node $\Omega = 0$ during first quadrature; for our purposes we leave $\Omega = 0$ as changes to the longitude do not affect the resulting light curves. $\hat{y}$ points ``up'', perpendicular to the projected motion of the planet in its orbit; it is the imaginary axis that defines the axis of reflection used when discussing the symmetry in the previous sub-section. The subscripts on the obliquity angles $\psi$ refer to changes of the spin axis orientation in the plane of the sky (``sky'') and along the line of sight (``LOS''). Following this definition, all references to planetary latitude and longitude hereafter are defined relative to the planet's rotation axis, rather than to the above Cartesian axes.  Therefore we have three independent variables that combine to give the components of our rotation vector: the rate $\omega_\mathrm{rot}$, and the two obliquity angles $\oblLOS$ and $\oblsky$ (see Figure \ref{fig:system_diagrams}). The rotation rate determines how much of the planet's area is visible, since at fast enough rotation significantly more than a single hemisphere can be visible at some point throughout secondary eclipse.

To get a sense of the timescales involved, we use two time parameters: the ratio of both the ``total'' eclipse duration ($T$ or $T_\mathrm{tot}$, commonly referred to as $t_1$--$t_4$) and either the ingress or egress duration ($\tau_\mathrm{i} =t_1$--$t_2$ or $\tau_\mathrm{e} = t_3$--$t_4$, equivalent for circular orbits), to the rotation period (see Figure \ref{fig:system_diagrams}). For $R_\mathrm{p} \ll R_\star$, $e=0$, and $b < 1$, these ratios are approximately \citep[see][for some of the reference definitions for eclipse durations]{Winn2010}
\begin{eqnarray}\label{eq:eclipse_durations}
    N_\mathrm{tot} &\equiv \frac{\Delta T_\mathrm{tot}}{P_\mathrm{rot}} \nonumber \\
    &\approx \frac{1}{\pi} \frac{P_\mathrm{orb}}{P_\mathrm{rot}} \left(\sqrt{1-b^2}\frac{R_\star}{a} + \frac{1}{\sqrt{1-b^2}}\frac{R_\mathrm{p}}{a} \right), \\
    N_\mathrm{i,e} &\equiv \frac{\tau_\mathrm{i,e}}{P_\mathrm{rot}} \nonumber \\
    &\approx \frac{1}{\pi} \frac{P_\mathrm{orb}}{P_\mathrm{rot}} \frac{1}{\sqrt{1-b^2}}\frac{R_\mathrm{p}}{a},
\end{eqnarray}
though for the actual calculations we solve for these durations numerically. These parameters represent the number of full rotations the planet makes across the entire secondary eclipse, and during just one of ingress or egress.

The actual rotation rates modeled are listed in Table \ref{table:warmJupiter_parameters}, which range from rotation to orbital rate ratios $\omega_\mathrm{rot}/\omega_\mathrm{orb} = 1$--300. This range is chosen to encompass the range of rates that would span the above eclipse durations. However, admittedly the high end of this range is likely to be out of range for many physical systems. For reference, using Equation \ref{eq:break-up_velocity}, a Jupiter-radius planet with density 1 g cm$^{-3}$ on a 10-day orbital period will have a break-up limit of $\sim 60$ rotations per orbit, that is $\omega_\mathrm{rot, max}/\omega_\mathrm{orb} = 60$. For HAT-P-18 b, which has a measured density of $\approx 0.25$ g cm$^{-3}$ and a 5.5-day orbital period, it may only rotate $\approx 16$ times per orbit before the break-up limit applies. With masses up to the deuterium-burning limit of $13 M_\mathrm{J}$ and an orbital period in the tens of days, we could set a very approximate maximum for our Warm Jupiters of $\sim 300$. However, we emphasize this is quite a generous upper limit, and we may expect a more realistic distribution of rotation rates to be well below this limit.

\subsection{Basis Variations with Rotation Rate Alone}\label{sec:eigenbases:rotation_rate}
Assuming zero obliquity, when $N_\mathrm{tot} < 1$, the eigenmaps are effectively restricted to one hemisphere. Otherwise, the planet will show different longitudes throughout an eclipse observation, and therefore the eclipse curve will encode variations from brightness structure across more of the planet's surface area than just a single hemisphere. If the rotation rate is known a priori, we can apply the Nyquist sampling argument to begin to understand the map at some spatial resolution. However, if the rotation rate is unknown there are multiple rotation rates that can produce similar eclipse curves. This degeneracy is unavoidable given the projection of a 2-D map to a 1-D time series, and has already been studied for quite some time in the context of orbital and rotation configurations \citep[see e.g.][]{Schwartz2016,Cowan2018}. Our aim is to understand how the ideal \emph{basis} for recovering a planet's map from its eclipse curve changes as a function of rotation state. While this exercise does not resolve the inherent degeneracies, by identifying and analyzing them, we can determine the level to which they may influence our observations in practice.

\begin{figure*}[htb!]
\begin{center}
\includegraphics[width=17cm]{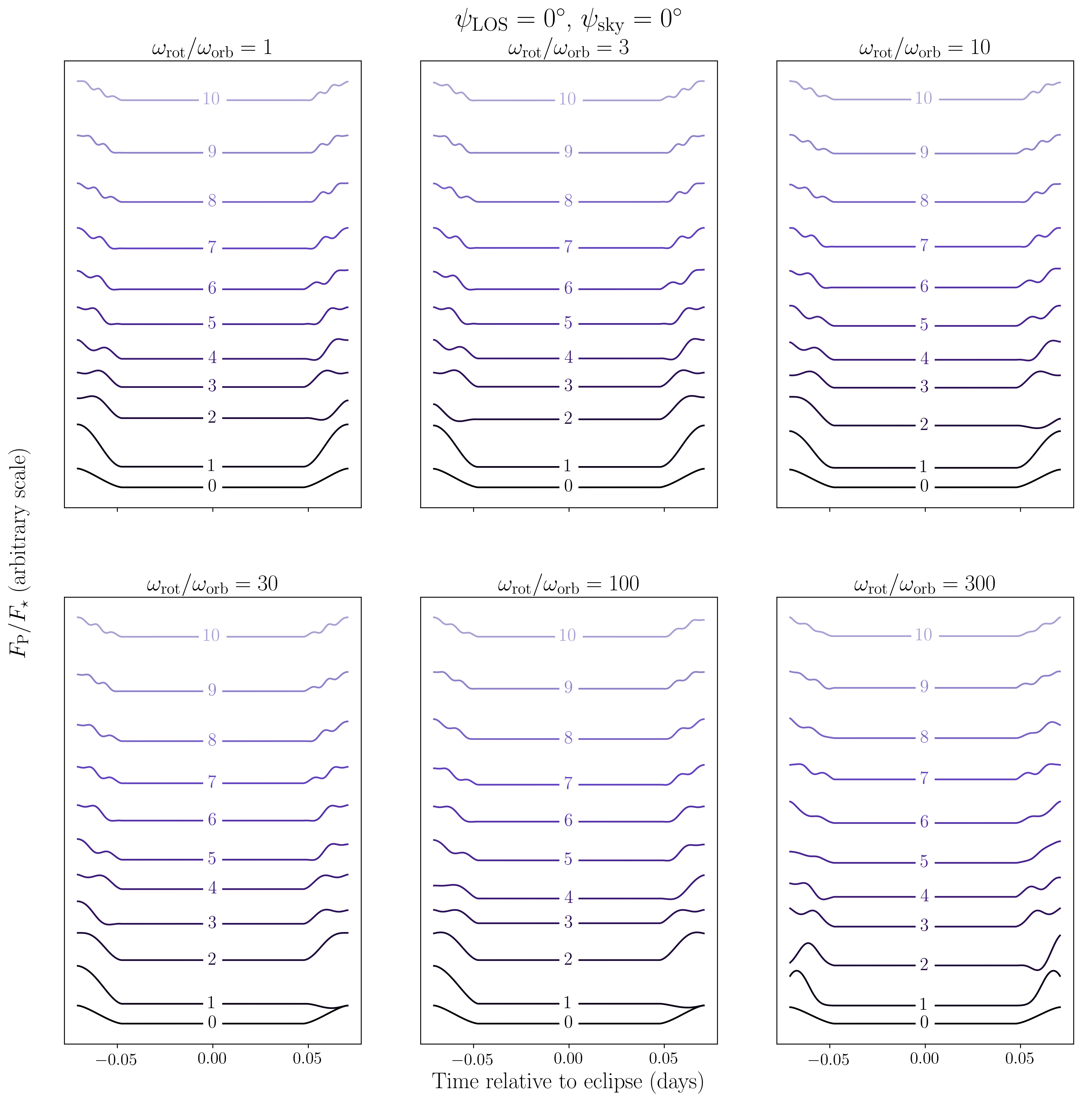}
\caption{The uniform brightness curve and the first 10 eigencurves with uniform component added for our Warm Jupiter model, at a range of rotation rates with zero obliquity. Rotation rates span the scale from ``slow'' , where the planet rotates a small fraction of a full rotation across eclipse ($N_\mathrm{tot} < 1$), to ``fast'' where the planet rotates through a considerable fraction of a full rotation. At the fastest rate the planet's rotation period approaches the duration of ingress or egress ($N_\mathrm{i, e} \rightarrow 1$). For reference, with our prototypical 10-day warm Jupiter $N_\mathrm{tot} = 1$ corresponds to $\omega_\mathrm{rot}/\omega_\mathrm{orb} \approx 71$, and $N_\mathrm{i, e} = 1$ corresponds to $\omega_\mathrm{rot}/\omega_\mathrm{orb} \approx 432$. One might notice that some eigencurves dip below the baseline where the planet is completely behind the star and contributes nothing to the observed flux. The eigencurves are initially generated from spherical harmonics, which each have zero net global brightness. Therefore, the eigencurves themselves start out as combinations of maps which each have zero net brightness. To ``reconsitute'' the eigencurves, we plot them here as additions on top of the uniform brightness curve. This allows for an eclipse curve composed of a single eigenmode to be unphysical, i.e.~ have net negative flux at some points.}
\label{fig:rotation-rate_eigencurves}
\end{center}
\end{figure*}

\begin{figure*}[htb!]
\begin{center}
\includegraphics[width=17cm]{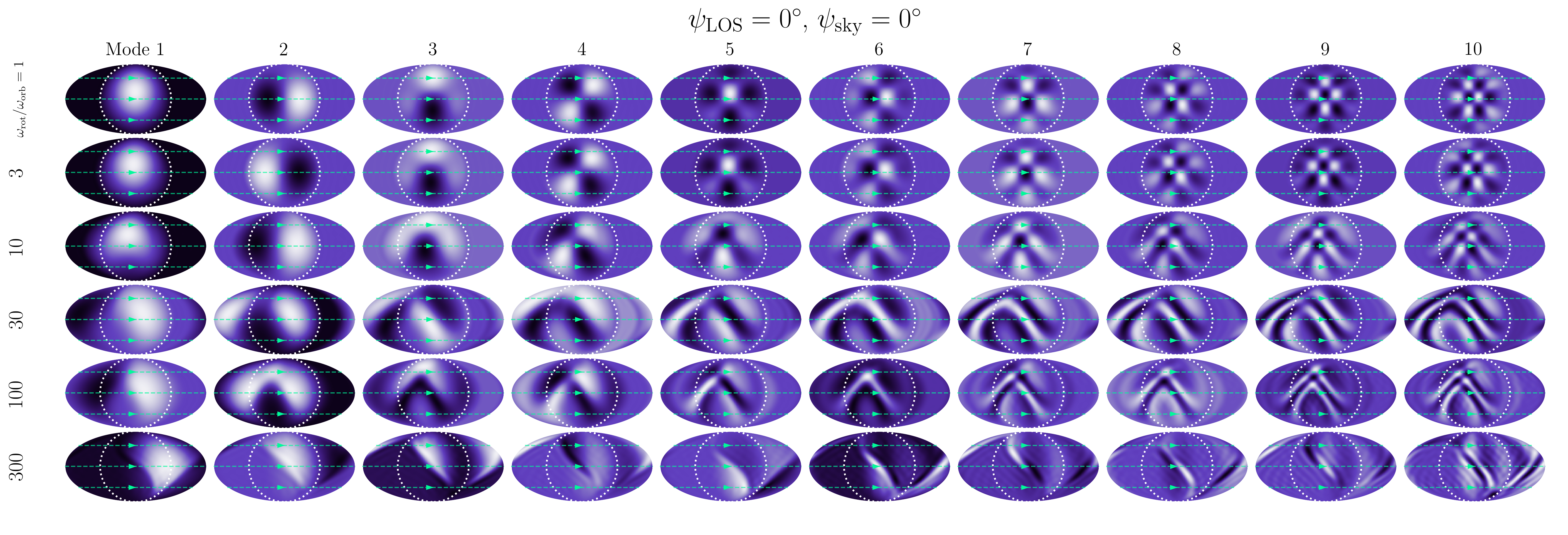}
\caption{The first ten modes of eigenmaps identified by PCA for the zero-obliquity Warm Jupiter case. We show the maps in a Mollweide projection that shows all longitudes and latitudes. The reference hemisphere, which is the hemisphere that faces the observer at the mid-point phase of ingress, is centered in each projection and outlined with a dotted white circle. Green dashed contours represent the equator and $\pm 45^\circ$ latitude lines, with arrows indicating the direction of rotation. Rotation rates span the scale from ``slow'' , where the planet rotates a small fraction of a full rotation across eclipse ($N_\mathrm{tot} < 1$), to ``fast'' where the planet rotates through a considerable fraction of a full rotation. At the fastest rate the planet's rotation period approaches the duration of ingress or egress ($N_\mathrm{i, e} \rightarrow 1$). For reference, with our prototypical 10-day warm Jupiter $N_\mathrm{tot} = 1$ corresponds to $\omega_\mathrm{rot}/\omega_\mathrm{orb} \approx 71$, and $N_\mathrm{i, e} = 1$ corresponds to $\omega_\mathrm{rot}/\omega_\mathrm{orb} \approx 432$.}
\label{fig:rotation-rate_eigenmaps}
\end{center}
\end{figure*}

Figure \ref{fig:rotation-rate_eigencurves} shows the effects of increasing the rotation rate on the eigencurve basis, with the corresponding eigenmaps in Figure \ref{fig:rotation-rate_eigenmaps}. The slowest rotation shown is identical to the synchronous case shown in Figure \ref{fig:base_eigenmodes}. As the rotation speeds up, the eigenmaps begin to stretch beyond a single hemisphere as more of the planet becomes visible. The regions of bright and dark begin to change shape: the boundaries spread out at roughly $45^\circ$ angles, and by the fourth row ($\omega_\mathrm{rot}/\omega_\mathrm{orb} = 30$) the structure is separating into two distinct patterns along these angles. The value of $45^\circ$ is close to the angle that the projected direction of the planet's orbit makes with the line tangent to where the planet appears to intersect the disk of the star (i.e.~the line tangent to the occulting edge). A map whose brightness gradient aligns with an occulting edge will produce the maximum variations in a light curve; the separation into two regions indicates that one region produces the majority of the light curve variations during ingress, and the other at egress. And since the occulting edges for ingress and egress are mirrored across the disk of the star, the angles of the regions are mirrored across the projection of that line onto the map. This gives a new perspective: if we slow the rotation back to synchronous, we see that the hemispherical maps are a superposition of the two mirror-angle patterns that separate at intermediate rotation rates. As the rotation quickens through the final row ($\omega_\mathrm{rot}/\omega_\mathrm{orb} = 300$), the features in each eigenmap become sharper: the maps are ``fine tuning'' to the shape of the occulting edge in order to maximize the brightness variations.

What we see overall is that there are 2 regimes of rotation rate in terms of the structure of their eigenmodes. The first is when rotation is close to synchronous, with a straightforward progression of increasingly finer structure across a single hemisphere yielding a monotonically increasing frequency of eclipse curve variations. Then, in the faster regime, where the rotation period is comparable to the time scales of eclipse, eclipse curves inform us about structure at two increasingly divergent regions of the globe, particularly structure whose brightness gradient aligns with the occulting edge.

\subsection{Basis Variations when Obliquity is Included}\label{sec:eigenbases:obliquity}
We now add in the effects of obliquity on top of the changes to rotation rate. Planetary obliquity further changes the map structures that are allowed to construct eclipse brightness variations. As we will show in this section, the trends in the shape of the eigencurves with rotation rate do not change appreciably when obliquity is added --- but those similar eigencurves are now produced by different eigenmaps. The slowest rotation rate in each case  are virtually identical to their zero-obliquity equivalents, since the effect of the rotation is minimal regardless of the direction the planet rotates. As rotation rate increases, the way in which the eigencurves change largely mimic the changes seen at zero obliquity in \S \ref{sec:eigenbases:rotation_rate}, but the eigenmaps adapt with shapes unique to their distinct rotation orientations. We consider two extreme cases of obliquity in the following sub-sections, with more intermediate cases available in Appendix \ref{sec:appendix}.

\subsubsection{Obliquity in the Plane of the Sky}\label{sec:eigenbases:obliquity:sky}
\begin{figure*}[htb!]
\begin{center}
\includegraphics[width=17cm]{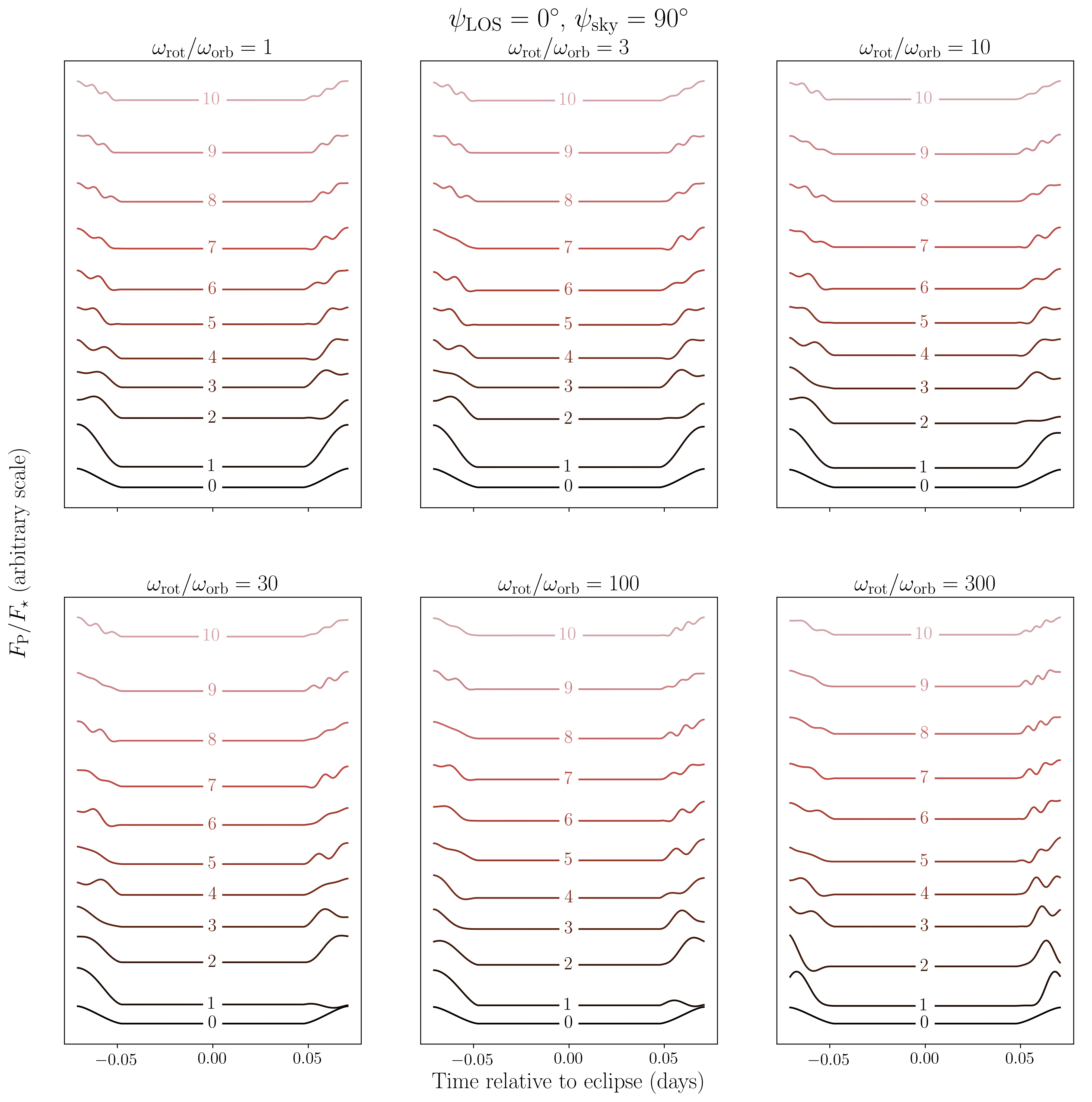}
\caption{The uniform brightness curve and the first 10 eigencurves with uniform component added for our Warm Jupiter model, at a range of rotation rates with $\oblsky = 90^\circ$. Rotation rates span the scale from ``slow'' , where the planet rotates a small fraction of a full rotation across eclipse ($N_\mathrm{tot} < 1$), to ``fast'' where the planet rotates through a considerable fraction of a full rotation. At the fastest rate the planet's rotation period approaches the duration of ingress or egress ($N_\mathrm{i, e} \rightarrow 1$). For reference, with our prototypical 10-day warm Jupiter $N_\mathrm{tot} = 1$ corresponds to $\omega_\mathrm{rot}/\omega_\mathrm{orb} \approx 71$, and $N_\mathrm{i, e} = 1$ corresponds to $\omega_\mathrm{rot}/\omega_\mathrm{orb} \approx 432$. One might notice that some eigencurves dip below the baseline where the planet is completely behind the star and contributes nothing to the observed flux. The eigencurves are initially generated from spherical harmonics, which each have zero net global brightness. Therefore, the eigencurves themselves start out as combinations of maps which each have zero net brightness. To ``reconsitute'' the eigencurves, we plot them here as additions on top of the uniform brightness curve. This allows for an eclipse curve composed of a single eigenmode to be unphysical, i.e.~ have net negative flux at some points.}
\label{fig:obl-sky90_eigencurves}
\end{center}
\end{figure*}

Changing the orientation of the planet's spin axis within the plane of the sky results in only subtle changes to the resulting eigenmodes, as shown for the limiting case of $\oblsky = 90^\circ$ (Figures \ref{fig:obl-sky90_eigencurves}--\ref{fig:obl-sky90_eigenmaps}). (The results for intermediate obliquities in the sky are shown in the Appendix, \S \ref{sec:appendix}, in Figures \ref{fig:obl-sky30_eigencurves}--\ref{fig:obl-sky60_eigenmaps}.) At the slowest rotation rates, regardless of how the spin axis is oriented, the curves and maps do not differ appreciably from the zero-obliquity case. However, when $N_\mathrm{tot} \sim 1$, we begin to see differences in the maps (Figure \ref{fig:obl-sky90_eigenmaps}). Now that the motion of the rotation and the apparent motion of the stellar limb across the planet are orthogonal, the primarily longitudinal orientation of the map structures becomes a more complicated mix of latitudinal and longitudinal variations. These changes allow for very similar eclipse curves, but demonstrate that at this range of rotation rates, planets with high sky-plane obliquities do not produce fundamentally different degeneracies than those encountered by planets without significant obliquities.

\begin{figure*}[htb!]
\begin{center}
\includegraphics[width=17cm]{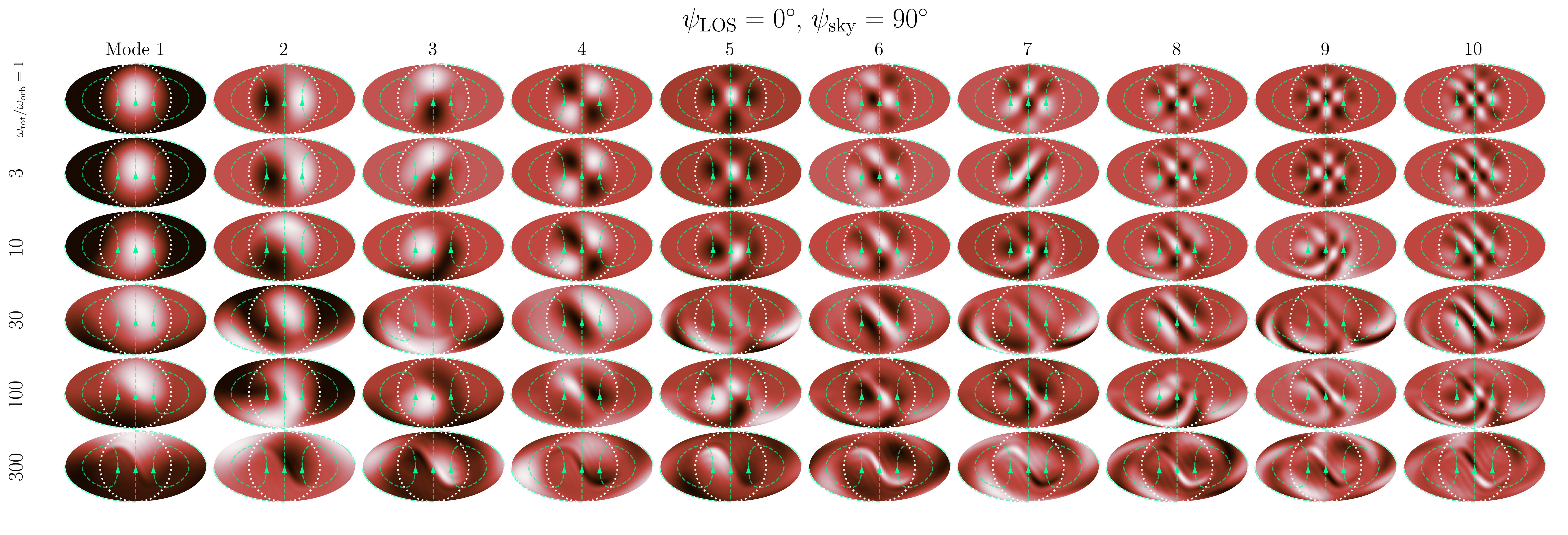}
\caption{The first ten modes of eigenmaps as identified by PCA for the Warm Jupiter case with $\oblsky = 90^\circ$. We show the maps in a Mollweide projection that shows all longitudes and latitudes. The reference hemisphere, which is the hemisphere that faces the observer at the mid-point phase of ingress, is centered in each projection and outlined with a dotted white circle. Green dashed contours represent the equator and $\pm 45^\circ$ latitude lines, with arrows indicating the direction of rotation. Rotation rates span the scale from ``slow'' , where the planet rotates a small fraction of a full rotation across eclipse ($N_\mathrm{tot} < 1$), to ``fast'' where the planet rotates through a considerable fraction of a full rotation. At the fastest rate the planet's rotation period approaches the duration of ingress or egress ($N_\mathrm{i, e} \rightarrow 1$). For reference, with our prototypical 10-day warm Jupiter $N_\mathrm{tot} = 1$ corresponds to $\omega_\mathrm{rot}/\omega_\mathrm{orb} \approx 71$, and $N_\mathrm{i, e} = 1$ corresponds to $\omega_\mathrm{rot}/\omega_\mathrm{orb} \approx 432$.}
\label{fig:obl-sky90_eigenmaps}
\end{center}
\end{figure*}

\subsubsection{Obliquity along the Line of Sight}\label{sec:eigenbases:obliquity:LOS}
\begin{figure*}[htb!]
\begin{center}
\includegraphics[width=17cm]{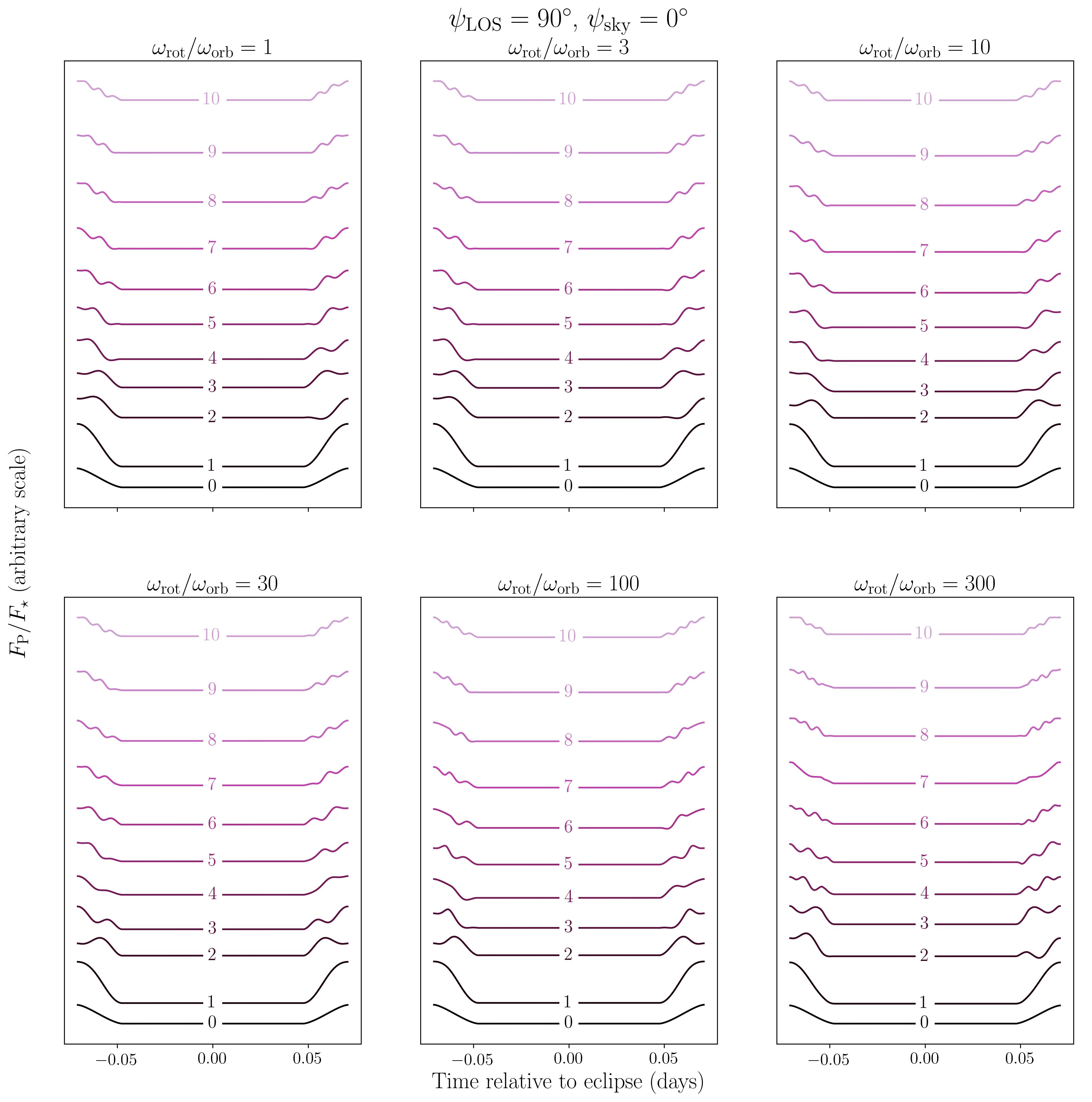}
\caption{The uniform brightness curve and the first 10 eigencurves with uniform component added for our Warm Jupiter model, at a range of rotation rates with $\oblLOS = 90^\circ$. Rotation rates span the scale from ``slow'' , where the planet rotates a small fraction of a full rotation across eclipse ($N_\mathrm{tot} < 1$), to ``fast'' where the planet rotates through a considerable fraction of a full rotation. At the fastest rate the planet's rotation period approaches the duration of ingress or egress ($N_\mathrm{i, e} \rightarrow 1$). For reference, with our prototypical 10-day warm Jupiter $N_\mathrm{tot} = 1$ corresponds to $\omega_\mathrm{rot}/\omega_\mathrm{orb} \approx 71$, and $N_\mathrm{i, e} = 1$ corresponds to $\omega_\mathrm{rot}/\omega_\mathrm{orb} \approx 432$. One might notice that some eigencurves dip below the baseline where the planet is completely behind the star and contributes nothing to the observed flux. The eigencurves are initially generated from spherical harmonics, which each have zero net global brightness. Therefore, the eigencurves themselves start out as combinations of maps which each have zero net brightness. To ``reconsitute'' the eigencurves, we plot them here as additions on top of the uniform brightness curve. This allows for an eclipse curve composed of a single eigenmode to be unphysical, i.e.~ have net negative flux at some points.}
\label{fig:obl-LOS90_eigencurves}
\end{center}
\end{figure*}

\begin{figure*}[htb!]
\begin{center}
\includegraphics[width=17cm]{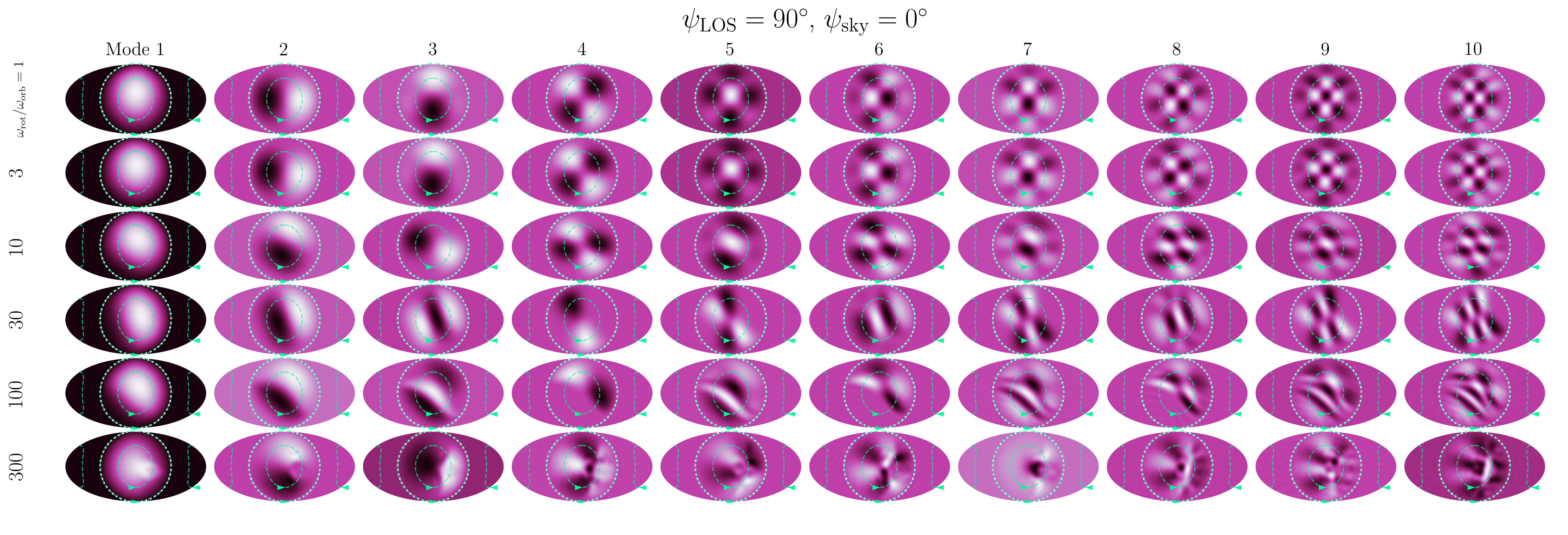}
\caption{The first ten modes of eigenmaps identified by PCA for the Warm Jupiter case with $\oblLOS = 90^\circ$. We show the maps in a Mollweide projection that shows all longitudes and latitudes. The reference hemisphere, which is the hemisphere that faces the observer at the mid-point phase of ingress, is centered in each projection and outlined with a dotted white circle. Green dashed contours represent the equator and $\pm 45^\circ$ latitude lines, with arrows indicating the direction of rotation. Rotation rates span the scale from ``slow'' , where the planet rotates a small fraction of a full rotation across eclipse ($N_\mathrm{tot} < 1$), to ``fast'' where the planet rotates through a considerable fraction of a full rotation. At the fastest rate the planet's rotation period approaches the duration of ingress or egress ($N_\mathrm{i, e} \rightarrow 1$). For reference, with our prototypical 10-day warm Jupiter $N_\mathrm{tot} = 1$ corresponds to $\omega_\mathrm{rot}/\omega_\mathrm{orb} \approx 71$, and $N_\mathrm{i, e} = 1$ corresponds to $\omega_\mathrm{rot}/\omega_\mathrm{orb} \approx 432$.}
\label{fig:obl-LOS90_eigenmaps}
\end{center}
\end{figure*}

As the obliquity along the line of the sight ($\oblLOS$) increases, there is an increasing fraction of the planet's map which will remain on the observer-facing hemisphere. In the extreme limit $\oblLOS \rightarrow \pm 90^\circ$, the observer stares down the rotation axis and only ever sees one hemisphere, regardless of how quickly the planet rotates. In order to demonstrate the effects of line-of-sight obliquity at their strongest, we show the eigenmodes for the limiting case $\oblLOS = 90^\circ$ in Figures \ref{fig:obl-LOS90_eigencurves}--\ref{fig:obl-LOS90_eigenmaps}. (The results for intermediate obliquities are available in the Appendix, \S \ref{sec:appendix}, in Figures \ref{fig:obl-LOS30_eigencurves}--\ref{fig:obl-LOS60_eigenmaps}.)

In this extreme case, the \emph{only} changes in brightness must come from the occultation --- otherwise, the map on the observable hemisphere will spin in place about its pole and the integrated flux will be constant. The qualitative changes to the eigencurves with increasing rotation rate are still similar to those of both the no-obliquity and sky-plane obliquity cases, (Figure \ref{fig:obl-LOS90_eigencurves}). At the fastest rotations, where $N_\mathrm{i,e} \sim 1$, the highest modes are a bit less regular and more ``noisy'': it is more difficult to generate high order variations with a rapidly rotating single hemisphere. The pole-on maps (Figure \ref{fig:obl-LOS90_eigenmaps}) adapt to increasing rotation rates by stretching out along one direction, creating a gradient for the occulting edge, but at even higher rotation rates the maps become very complicated structures which do not have a simple intuitive explanation, other than these are the maps computed to have the strongest gradients. As with the sky-plane obliquities, there are theoretical maps which yield a basis of curves at a range of frequencies, in a way that is not fundamentally different from the sky-plane obliquity and zero obliquity cases. However, it is important to note that this makes no assumptions about it truly \emph{plausible} brightness structures on an actual planet --- only that these are the structures that would in principle drive the strongest signals. It is interesting from a mathematical perspective to dive into the peculiar changes to the information content with the most extreme obliquity cases, but we emphasize them here primarily to show the structures particular to those extrema of possible spin geometries, with more moderate obliquities demonstrating some mixture of the geometric effects at zero and maximum obliquity, as shown in the figures included in the Appendix.

\subsection{Basis Variations with Obliquity Alone}\label{sec:eigenbases:only-obliquity}
We now take one step back, to consider the effects of obliquity alone, as in the effects of orientation but not rotation rate specifically. We consider a planet whose map has no longitudinal dependence: this is a map composed entirely of zonal ($m=0$) harmonics. Our knowledge from the previous sub-sections is that if the planet's map has brightness gradients aligned with the occulting edge, it can drive brightness variations within ingress and/or egress. Since purely zonal maps are insensitive to rotation, the gradient must come from latitudinal gradients on the map. For planets with zero obliquity and zero impact parameter, latitudinal brightness variations are indistinguishable from a uniform map at equivalent disk-integrated brightness. The projected arc of the occulting edge onto the planet disk exhibits latitudinal symmetry. This symmetry can be broken in two ways, the first coming from a non-zero impact parameter ($0<b<1$). In this case we break the latitudinal symmetry: the portions of the arc that occult the northern and southern hemispheres are now aligned differently and therefore shorten or lengthen the duration of eclipse accordingly depending on the amount of north-south asymmetry. This is a well-established outcome of planetary mapping \citep[see e.g.][]{Rauscher2007b,dew12,maj12,Cowan2018}. The other way to probe latitudinal structure is to tilt the planet. From the observer's perspective, this allows some of the latitudinal variations to mimic the behavior of longitudinal structure for zero-obliquity planets. As we will see, there are two types of changes to the uniform light curve from these oblique latitudinal structures: symmetric and asymmetric, where the symmetry in question is reflective about the midpoint of eclipse.

\subsubsection{Obliquity Alone in the Plane of the Sky}\label{sec:eigenbases:only-obliquity:sky}
\begin{figure*}[htb!]
\begin{center}
\includegraphics[width=17cm]{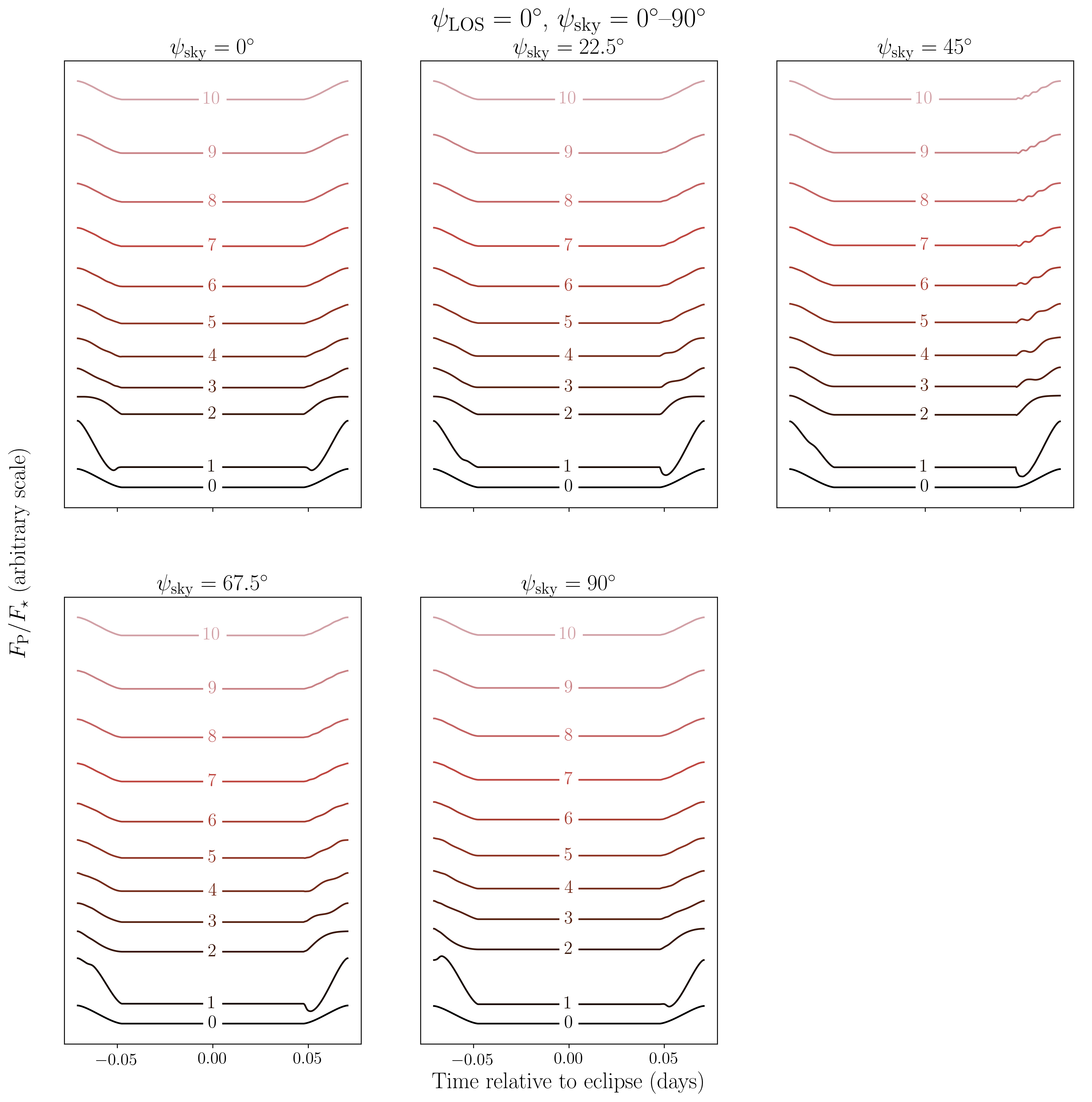}
\caption{The uniform brightness curve and the first 10 eigencurves with uniform component added for our Warm Jupiter model, at a range of obliquities in the sky plane ($\oblsky$). The maps shown here are purely latitudinal, i.e.~are insensitive to rotation rate, to show the effects of axial orientation alone. One might notice that some eigencurves dip below the baseline where the planet is completely behind the star and contributes nothing to the observed flux. The eigencurves are initially generated from spherical harmonics, which each have zero net global brightness. Therefore, the eigencurves themselves start out as combinations of maps which each have zero net brightness. To ``reconsitute'' the eigencurves, we plot them here as additions on top of the uniform brightness curve. This allows for an eclipse curve composed of a single eigenmode to be unphysical, i.e.~ have net negative flux at some points.}
\label{fig:obl-sky_zonal_eigencurves}
\end{center}
\end{figure*}

\begin{figure*}[htb!]
\begin{center}
\includegraphics[width=17cm]{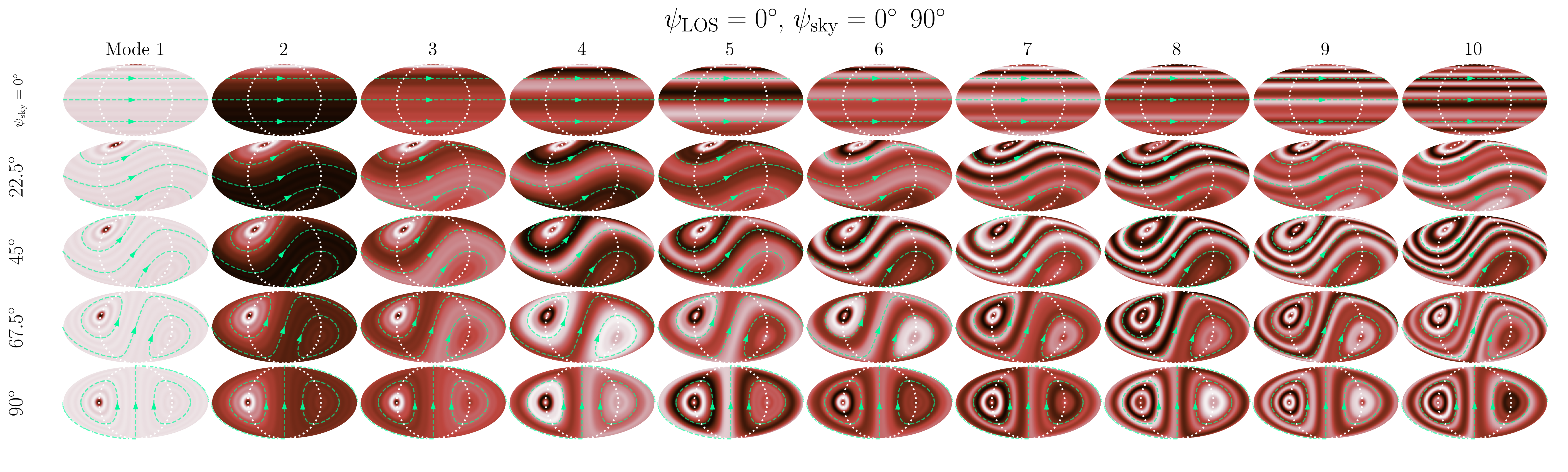}
\caption{The first ten modes of eigenmaps as identified by PCA for the Warm Jupiter case, at a range of obliquities in the sky plane ($\oblsky$). The maps shown here are purely latitudinal, i.e.~are insensitive to rotation rate, to show the effects of axial orientation alone. We show the maps in a Mollweide projection that shows all longitudes and latitudes. The reference hemisphere, which is the hemisphere that faces the observer at the mid-point phase of ingress, is centered in each projection and outlined with a dotted white circle. Green dashed contours represent the equator and $\pm 45^\circ$ latitude lines, with arrows indicating the direction of rotation.}
\label{fig:obl-sky_zonal_eigenmaps}
\end{center}
\end{figure*}

Our first case is for sky-plane obliquity (Figures \ref{fig:obl-sky_zonal_eigencurves} and \ref{fig:obl-sky_zonal_eigenmaps}), where we see two major differences from the eigenmodes where longitudinal variations are allowed. The first is that only the first couple of eigencurves have appreciable structure, with modes $\geq 5$ only negligibly different from the uniform light curve in all except the $\oblsky=45^\circ$ case. Second is that as one moves to greater $\oblsky$ angles, the lowest non-uniform eigencurves show asymmetry.

The first eigenmap shows a bright point centered at each planet's north pole; this makes sense as a sharp bright feature is able to create a sharp feature in the light curve. Moving to the second mode eigenmaps, a slightly more extended polar spot, we see that in the corresponding eigencurves the asymmetry between ingress and egress increases up to $\oblsky = 45^\circ$ and then remains through the maximum obliquity. At the lowest modes we are picking up mostly hemispherical asymmetries. Imagine reflecting the planet map about $\hat{y}$, the axis perpendicular to the projected orbital motion. As obliquity increases for a longitudinally-symmetric map, a hemispherical asymmetry can contribute more and more to breaking reflective symmetry across the midpoint of eclipse. This is effectively allowing the latitudinal bands to mimic what would be longitudinal structure at low obliquity.

At higher modes, higher values of $\oblsky$ alone are not sufficient to produce observable structure. Now it matters how closely the obliquity angle is aligned with the normal of the occulting edge. In this case, the planet's map cannot produce gradients at two different angles to accommodate ingress and egress, as it could with the eigenmaps in previous cases. At $\oblsky = 45^\circ$ its features are well aligned with the ingress occulting edge, but almost perfectly orthogonal to the egress occulting edge, yielding almost no variations during egress. We see in the eigencurves of the planet with $\oblsky = 45^\circ$ that it maintains the most discernible variations, again preferentially during ingress. 

\subsubsection{Obliquity Alone along the Line of Sight}\label{sec:eigenbases:only-obliquity:LOS}
The behavior of the latitudinal eigenmodes for planets with obliquity along the line of sight is similar to that of obliquity in the plane of the sky, in that in increasing mode number we start with sharp polar bright spots that then diffuse out from the pole. It is also similarly difficult to generate appreciable light curve variations beyond just a few modes. At non-zero values of $\oblLOS$ the second eigencurve shows a ``shortening'' of the eclipse dip which comes from the latitudinal contrast. The asymmetries seen in the cases with $\oblsky$ are now absent, as the features in the $\oblLOS$ eigenmaps are now completely symmetric across $\hat{y}$ (as defined in \S \ref{sec:eigenbases}).

Taken together, these show the effects of latitudinal map variations as we extend to different rotation states. The impact parameter matters, as this sets the angle of the occulting edge relative to the planet's orbit and therefore how well latitudinal gradients can drive eclipse curve variations. Obliquity in the plane of the sky maximizes asymmetry between the shapes of ingress and egress, and especially for eclipse sub-structure when the latitudinal gradient aligns with the normal to the occulting edge, i.e.~where for impact parameter $b$
\begin{equation}\label{eq:maximum_gradient}
    \cos\oblsky = \cos b.
\end{equation}

\begin{figure*}[htb!]
\begin{center}
\includegraphics[width=17cm]{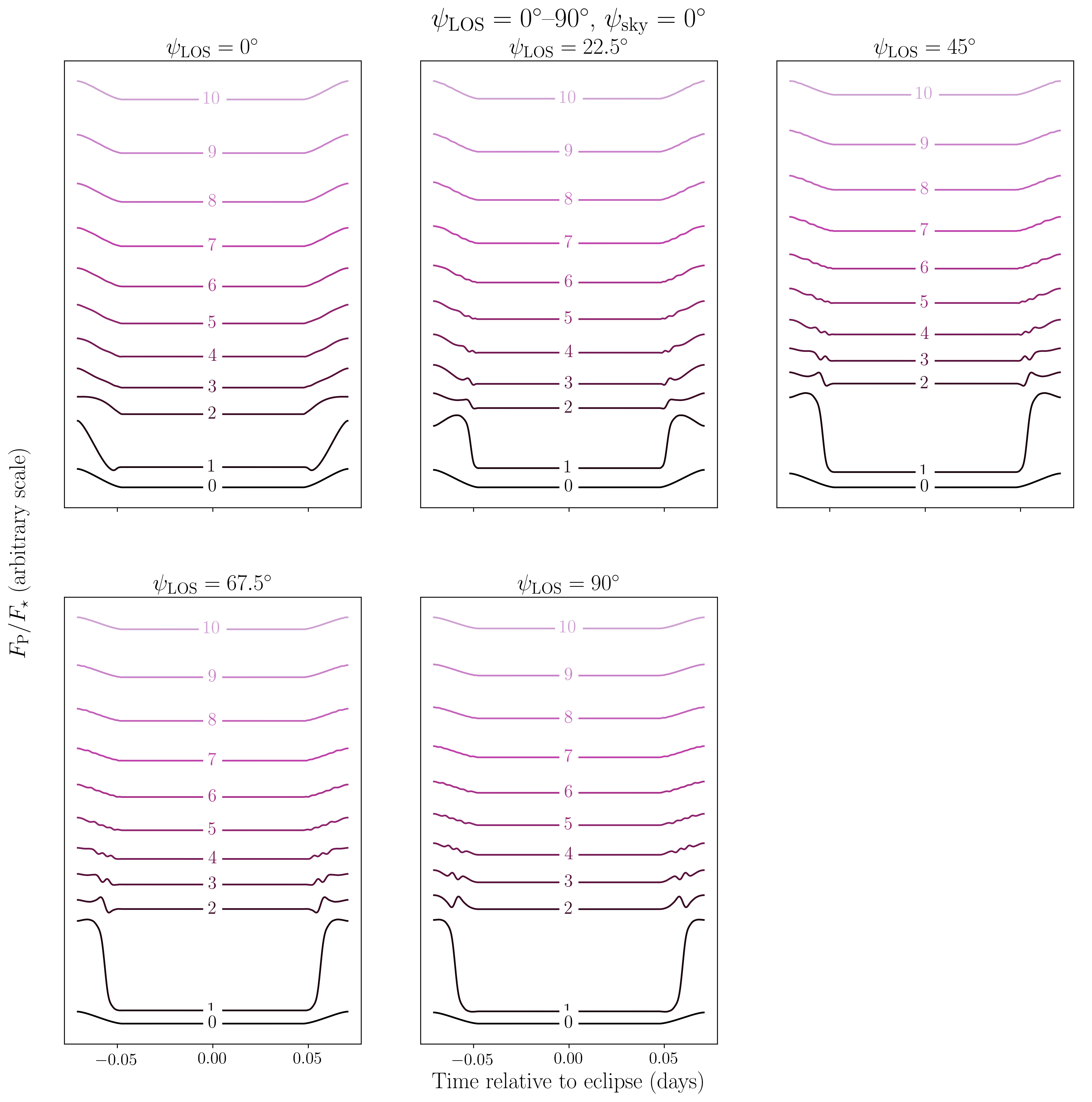}
\caption{The uniform brightness curve and the first 10 eigencurves with uniform component added for our Warm Jupiter model, at a range of obliquities along the line of sight ($\oblLOS$). The maps shown here are purely latitudinal, i.e.~are insensitive to rotation rate, to show the effects of axial orientation alone. One might notice that some eigencurves dip below the baseline where the planet is completely behind the star and contributes nothing to the observed flux. The eigencurves are initially generated from spherical harmonics, which each have zero net global brightness. Therefore, the eigencurves themselves start out as combinations of maps which each have zero net brightness. To ``reconsitute'' the eigencurves, we plot them here as additions on top of the uniform brightness curve. This allows for an eclipse curve composed of a single eigenmode to be unphysical, i.e.~ have net negative flux at some points.}
\label{fig:obl-LOS_zonal_eigencurves}
\end{center}
\end{figure*}

\begin{figure*}[htb!]
\begin{center}
\includegraphics[width=17cm]{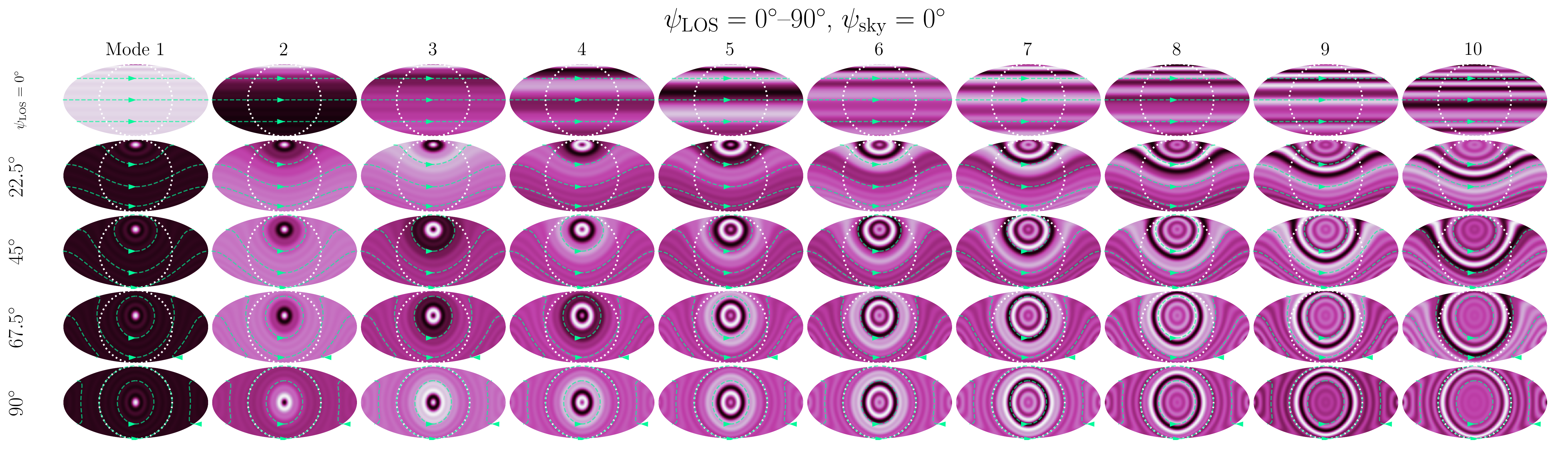}
\caption{The first ten modes of eigenmaps as identified by PCA for the Warm Jupiter case, at a range of obliquities along the line of sight ($\oblLOS$). The maps shown here are purely latitudinal, i.e.~are insensitive to rotation rate, to show the effects of axial orientation alone. We show the maps in a Mollweide projection that shows all longitudes and latitudes. The reference hemisphere, which is the hemisphere that faces the observer at the mid-point phase of ingress, is centered in each projection and outlined with a dotted white circle. Green dashed contours represent the equator and $\pm 45^\circ$ latitude lines, with arrows indicating the direction of rotation.}
\label{fig:obl-LOS_zonal_eigenmaps}
\end{center}
\end{figure*}

\subsection{A Summary of Eclipse Curve Structure with Rotation}\label{sec:eigenbases:summary}
Taken together, the eigenbases in the previous section paint a picture where arbitrary eclipse curve structures can be generated across a large range of rotation rates and obliquities. The modes of each eigenbasis represent the brightness patterns which are the most ``mappable'', that is provide the most signal. Most of the eigencurves resemble sinusoidal variations, mimicking the orthonormality of sinusoids ordered by frequency for a 1-dimensional dataset. At the simplest case, with slow rotation and zero obliquity, the maps have initial dipole shapes in longitude and latitude that then sub-divide into checkerboard patterns of light and dark through successive modes. The path of the star's disk across the planet's observed disk sweeps across a sinusoidal gradient of light and dark that adapts to the reflected symmetry of the occultation between ingress and egress. As rotation rate increases, there eventually become two distinct regions of the planet's map that produce variations in ingress and egress separately. As obliquity changes, the interplay between the direction of rotation and the path of the occulting edge of the stellar disk produce more complicated eigenmaps, but the machinery of PCA applies the same principle of using brightness gradients to impart structure. As a special case, when only latitudinal map structure exists, the rotation rate will not change the observed brightness map --- this is essentially a way of isolating the effects from planetary obliquity alone. Obliquity in the plane of the sky determines the orientation of the planet's latitudinal bands relative to the occulting edges and therefore influences how strong of a gradient can be imparted into the eclipse curve. Since gradients drive structure, this directly determines how much one can learn about the planet's map through eclipse mapping, and suggests that planets at particular obliquities $oblsky$ will be more ``mappable'' than others.

\section{Retrieving Maps when Rotation is Unknown}\label{sec:retrieval}
\begin{deluxetable}{lc}[htb!]
\tabletypesize{\footnotesize}
\tablewidth{0pt}
\tablecaption{Mock observational parameters for the HAT-P-18 system}
\tablehead{\colhead{Parameter} & \colhead{HAT-P-18 b}}
\startdata
Integration Time (s) & 0.55 \\
Signal-to-Noise Ratio & $\approx 500$ \\
Total Eclipse Duration (T$_1$--T$_4$, days) & 0.119 \\
Ingress/Egress Duration (T$_1$--T$_2 =$ T$_3$--T$_4$, days) & 0.016 \\
Partial Phase Duration ($\Delta T / P_\mathrm{orb}$) & 0.2\tablenotemark{a} \\
Partial Phase Duration (days) & 1.10\tablenotemark{a} \\
\enddata
\tablenotetext{a}{Partial phase observations are not included in the original example science program.}
\label{table:HAT-P-18b_parameters}
\end{deluxetable}

We have seen in \S \ref{sec:eigenbases} that the rotation state (speed and obliquity) of a planet directly affects what brightness structures impart the highest amplitude structures in an eclipse curve. Where similar eclipse curve structures exist from two planets with different maps and rotation states, there is ambiguity. This claim is theoretical, as we have intentionally made no assumptions about what sorts of map structures are more likely to occur on real planets. Also, in pursuit of understanding the theoretical geometrical limits to the information available through eclipse, we have assumed an artificially high time resolution in our model eigencurves, with none of the noise present in real observations. It would be helpful to understand how much of a problem these limitations will be in the near- to medium-term. To make a small step in tackling this problem, we construct model light curves with the system and JWST observational parameters\footnote{This mock observation is based on the Example Science Program \#29; its observational parameters can be found at \url{https://jwst-docs.stsci.edu/jwst-near-infrared-camera/nircam-example-science-programs/nircam-time-series-imaging-of-hat-p-18-b}.} of HAT-P-18, which is a K2V host to a 1 $R_\mathrm{J}$, 0.2 $M_\mathrm{J}$ planet at an orbital period of $\approx 5.5$ days ($\approx 0.05$ AU). While this is a shorter orbital period than the model Warm Jupiter we used in previous sections, the host's cooler-than-solar spectral type means that the planet's equilibrium temperature is $\approx 850$ K \citep{Hartman2011}, earning it the classification of a modestly ``warm'' giant planet. We sample the eclipse curve at 200 data points evenly spaced in time, with a bin width of approximately 51 seconds. We do not add any noise to the time dimensions, though the expected timing error for JWST data is expected to be about an order of magnitude smaller than our bins. We also make a number of simplifying assumptions in the interest of constraining the scope of this work:
\begin{enumerate}
    \item Our simulated data of HAT-P-18 b are generated from various hotspot map models. Realistically speaking, a hotspot near the equator or mid-latitudes is only a physically motivated map for slow rotations, since fast enough rotation is expected to smear out any significant longitudinal variations. We use it as an example of a simple map structure and choose to apply it across all modeled rotation rates for consistency. One can consider this a case of a basic upper limit to the signal of a low-complexity map, through a hotspot which will cycle in and out of view through rotation.
    \item the planet has zero eccentricity.\footnote{The non-zero eccentricity of the real HAT-P-18 b demonstrates that tides have not had time to circularize its orbit, potentially indicating that they also have not synchronized and aligned its rotation axis. Our choice to exclude eccentricity is merely a simplification; eclipse mapping is otherwise applicable to planets with significant non-zero orbital eccentricities.}
    \item we are able to observe with a signal-to-noise greater than the model observations in the example program. The program covers 3 successive eclipses which provides a signal-to-noise of $\approx 500$ per integration time across the NIRCam F210M and F444W filters. This signal, while sufficient to capture the eclipse depth, is not quite sufficient to capture map structure within ingress or egress (even assuming zero correlated detector systematics). Therefore, we show simulations of eclipse curves with precisions that would be feasible in $\sim 10$ orbits, representing an artificial increase of $\sim$2 in the signal-to-noise. This was chosen as a rough lower limit to where confusion in the map structure may result from an inaccurate rotation rate, as seen in the following sub-sections.
    \item Additionally, we are assuming no orbit-to-orbit variation in the observed maps --- so when numbers of orbits are mentioned in this paper, it is primarily as a proxy for signal-to-noise rather than a realistic simulation of observations of multiple successive eclipses.
    \item Finally, we do not account for any tidal or rotational deformations that would induce a non-spherical shape for the planet. For rotation rates approaching the theoretical break-up limit, one might expect planets to be stretched equatorially, and this stretching could change the resulting frequency of eclipse curve variations imposed by longitudinal structure.
\end{enumerate}




The hotspot has an intensity given by
    \begin{equation}\label{eq:hotspot_function}
F\left(\phi,\theta\right) \propto \exp\left[\sigma_{\theta}^{-1/2} \cos\!\left(\theta-\theta_0\right) + \sigma_{\phi}^{-1/2} \sin\theta \sin\theta_0 \cos\!\left(\phi-\phi_0\right) \right]
\end{equation}
where $\left(\phi_0, \theta_0\right)$ is the position of the hotspot's center and $\sigma_\phi$ and $\sigma_\theta$ represent the angular scales of the hotspot in each dimension. In practice, STARRY automatically normalizes the global flux to the assumed planet-to-star luminosity ratio after the map is generated. For all cases shown here we also assume symmetric hotspots ($\sigma_\phi = \sigma_\theta$).

We then ``retrieve'' the maps by fitting the eclipse curves --- generated from a STARRY model with a given rotation state --- with a set of eigencurves where we do not assume we know the rotation state used to generate the data. The data are generated from planet models rotating from synchronous up in half orders of magnitude to the rough order of the break-up limit of HAT-P-18 b ($\omega_\mathrm{rot}/\omega_\mathrm{orb}=30$ versus about 16 for the break-up limit, see Table \ref{table:warmJupiter_parameters}). We limit our fitting exercise to a coarse grid matching the half orders in the rotation rates used to generate the data, that is assuming we have an uninformed prior on rotation rate that is only limited by the theoretical upper limit from break-up. This demonstrates the outcomes of our fitting routine when we happen to ``tune'' our rotation frequency to a more or less accurate value.

We show retrievals for three example cases. Given the conclusions outlined in \ref{sec:eigenbases:summary}, we would like to show cases which demonstrate the findings that rotation rate and obliquity can influence what map we infer for a given observed eclipse curve. Since longitudinal dipoles are typically found as the principal modes of most of the eigenbases, an equator-centered hotspot is a first choice for a simulated planet. One case focuses on the effects of rotation rate, another the additional contribution of obliquity to variable rotation rate. One final case shows how one might fit a longitudinally-symmetric planet map, as might be expected to first order for a planet with fast enough rotation, where the finding was that the value of $\oblsky$ should have the greatest effect in the signal retrievable in eclipse. In more detail, the 3 cases include:

\begin{enumerate}
    \item A planet with zero obliquity, whose hotspot is centered on the equator and 30 degrees east of the longitude that is sub-observer at the time of mid-ingress. Within this scenario we show two sub-cases:
    \begin{enumerate}
        \item (Figure \ref{fig:retrieved_zero_obliquity}) the ``base'' case, where the data is limited to secondary eclipse, and
        \item (Figure \ref{fig:retrieved_zero_obliquity_partial-phase}) a ``partial phase'' case when  we include $\pm$10\% of the orbital period on either side of secondary eclipse.  As the eclipse duration is $\sim$2\% of the orbital period, this amounts to an order of magnitude more observing time. This is to show an example of how additional information can place a strong prior on a fit to an eclipse curve. We also use 200 data points here, with our bins spanning approximately 40 minutes each --- therefore, we are looking at the rotational information from the phase curve rather than spatial information from the eclipse.
    \end{enumerate}
    \item (Figure \ref{fig:retrieved_oblLOS45-oblsky45}) The same map in the planet's coordinates as above, but where the planet's rotation axis has a $45^\circ$ angle both along the line of sight as well as away from the plane of the sky ($\oblLOS=\oblsky=45^\circ$).\footnote{Partial phase results are not shown for this case. In our simulations the difference in results from the zero-obliquity partial-phase case were negligible.} In this case we model different rotation rates, but still assume zero obliquity for our eigenbasis.
    \item (Figure \ref{fig:retrieved_only-obliquity}) Finally, a polar hotspot, centered at $+90^\circ$ latitude, with the same angular extent as above. In this case, instead of varying the rotation rate (as a polar hotspot is symmetric with respect to longitude), we vary the sky-plane obliquity ($\oblsky$) between 0--90$^\circ$. We do not limit our eigenbases to only longitudinal symmetry, as was done in \S \ref{sec:eigenbases:only-obliquity} as a demonstration; only the map that generate the data they fit is longitudinally symmetric.
\end{enumerate}

We employ the least-squares solver in Numpy to solve for the best fit using a successively larger number of eigencurves, estimating an appropriate number of eclipse curves (parameters) to include in our model via a Bayesian Information Criterion (BIC),
\begin{equation}
    \mathrm{BIC} = n_\mathrm{par} \ln\!\left(n_\mathrm{data}\right) - 2\ln\!\left(\mathcal{L}\right)
\end{equation}
where $n_\mathrm{par}$ is the number of parameters (eigencurves), $n_\mathrm{data}$ is the number of data points, and $\ln\!\left(\mathcal{L}\right)$ is the log likelihood of the model compared with the data. Then, one can argue the model with the lowest BIC value is preferred and therefore should represent the best estimate of the number of parameters\footnote{A common heuristic is to argue that, in comparisons between a model with fewer parameters and one with a greater number, one only has sufficient evidence to prefer the model with more parameters if the BIC is reduced by some minimum number, such as 2 or 6. In our analysis we simply show the distributions of models with minimum BIC values.}. We generate 10000 realizations of observations by applying a Gaussian noise profile to the model planet data, then run our fitting routine. This yields a distribution of the eigencurve coefficients, a distribution of the appropriate number of eigencurves warranted by the data, and a distribution of observed light curves and inferred brightness maps. We show these and discuss our interpretations in the following sub-sections.

\subsection{Retrieving Zero-Obliquity Maps}\label{sec:retrieval:zero-obliquity}
\begin{figure*}[htb!]
\begin{center}
\begin{tabular}{c}
\includegraphics[height=4cm]{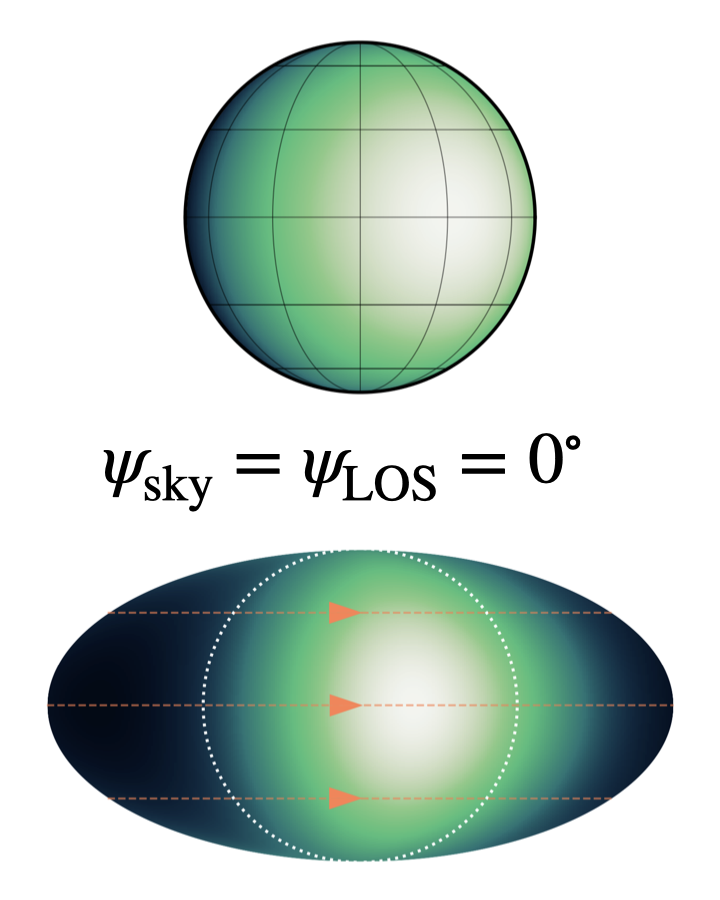} \\
\includegraphics[width=17cm]{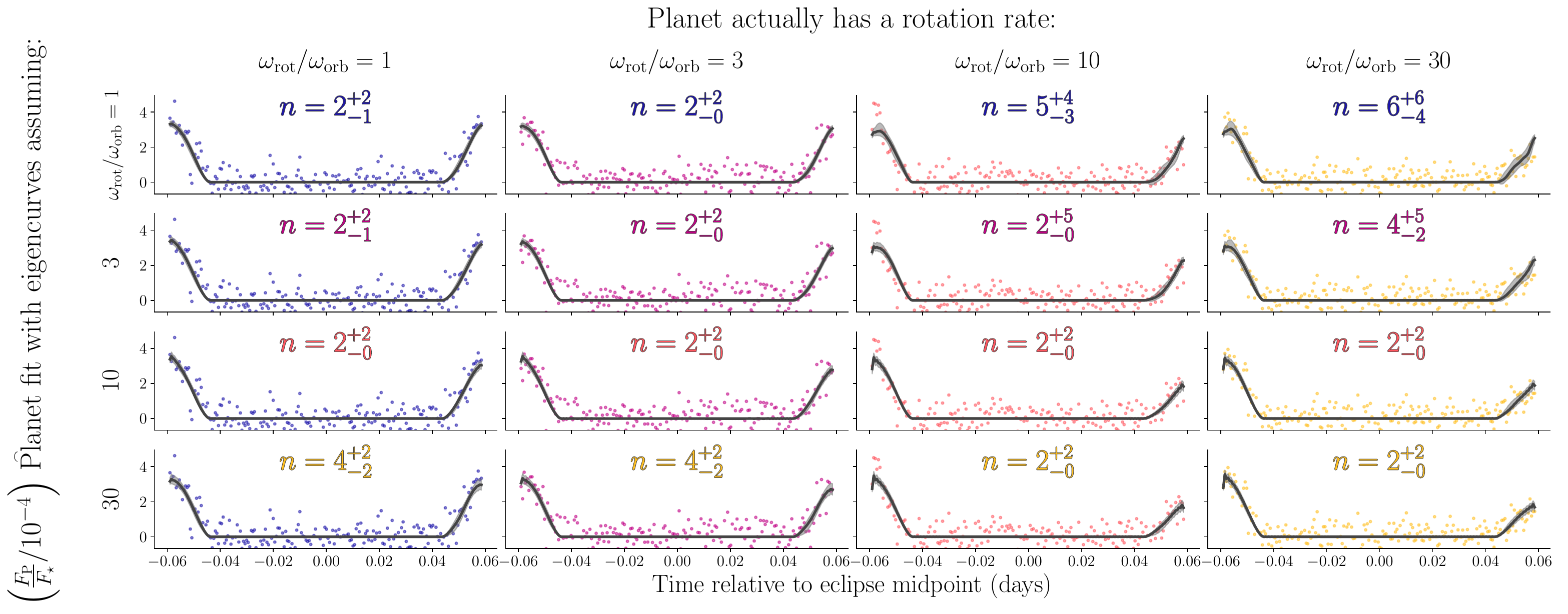} \\
\includegraphics[width=12cm]{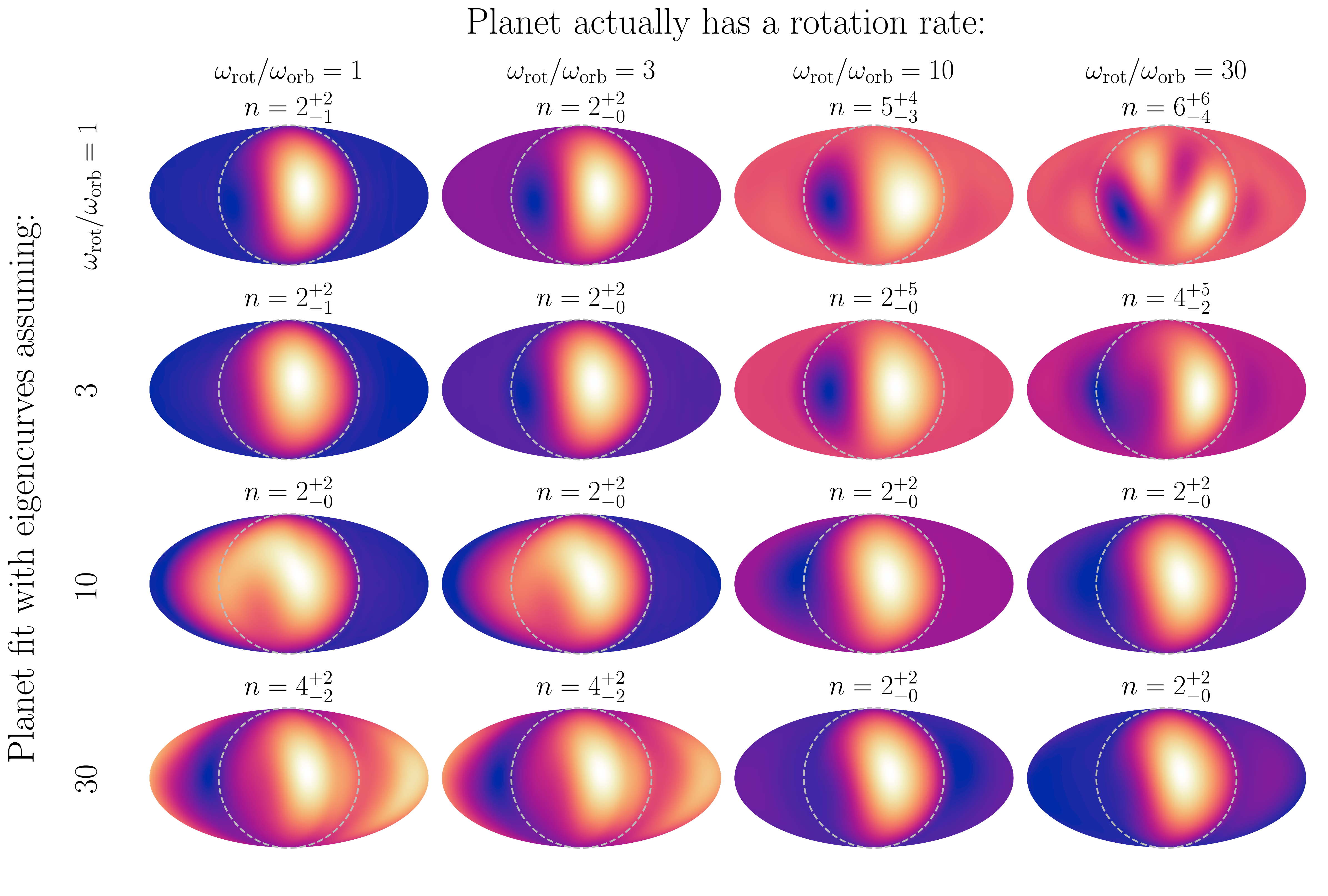} \\
\end{tabular}
\caption{Eclipse maps and curves from fits to simulated data from model planets rotating at a range of rates $\omega_\mathrm{rot}/\omega_\mathrm{orb} = 1$--30. The input map is shown at the top in green, as it would be oriented to the observer at the time of mid-ingress. The equivalent full-globe Mollweide projection is shown just below to compare with the equivalent projections of the retrieved maps. $\omega_\mathrm{rot}/\omega_\mathrm{orb}$ is equivalent to how many times the planet rotates per orbit. We generate 10000 realizations of Gaussian noise for each simulated eclipse curve, and fit each with some number $n$ of eigencurves (i.e.~parameters), where $n$ is determined by a Bayesian Information Criterion (BIC). The mode and $\pm 1 \sigma$ of $n$ are shown for each case. Recovered curves and maps are shown based on the rotation rate of the model used to generate the data ($x$-axis), versus the rotation rate of the eigenbasis used to fit the data ($y$-axis). Each sub-plot of the eclipse curve grid shows the range of uncertainty used to artificially noise the simulated data, with one example realization of the 10000 in colored points. The fit is shown in greyscale; the solid line is the median of all fits and a shaded region shows the $\pm 1 \sigma$ ($\approx$16th and 84th percentiles) range of fits. We show maps generated using the median values of the coefficients from each eigenmap.}
\label{fig:retrieved_zero_obliquity}
\end{center}
\end{figure*}

The input hotspot map and resulting eclipse maps and corresponding curves are shown in Figure \ref{fig:retrieved_zero_obliquity}. The diagonal in each grid of retrieved curves and maps represents the cases in our retrieval where the model eigencurves are tuned to the correct rotation rate. In these cases we expect the model to be able to capture the structure of the eclipse curves with just a few eigencurves, and indeed each of these fits has a distribution of eigencurves with a median of 2. The top row of the figures show the fits with an eigenbasis that assumes synchronous rotation. In this scenario, at low true rotation rates the simple hotspot structure is recovered, but for the synchronous eigenbasis fitting the data of fastest rotation, the fits on average need more eigencurves to fit the sharper turn-over of the eclipse curve during ingress. To reproduce this curvature, a more complex hemispherical brightness map is needed: the retrieved median map has its brightest region near the true location of the hotspot, but also adds an additional brightness gradient in order to capture the asymmetry observed between ingress and egress. Recall that we essentially called our coarse, 4-point grid in rotation rate representative of a uniform prior in rotation rate. We can roughly put a constraint on the rotation rate from this very coarse grid search if we prefer eigenbases which require the fewest eigencurves to fit the data. In the case of the planet rotating at the maximum of $\omega_\mathrm{rot}/\omega_\mathrm{orb} = 30$, we would find that, within $\sim 1/2$ an order of magnitude, eigenbases generated near the correct rotation rate on average use fewer eigencurves to fit the data.

\begin{figure*}[htb!]
\begin{center}
\begin{tabular}{c}
\includegraphics[height=4cm]{zero-obliquity_inputmaps.png} \\
\includegraphics[width=17cm]{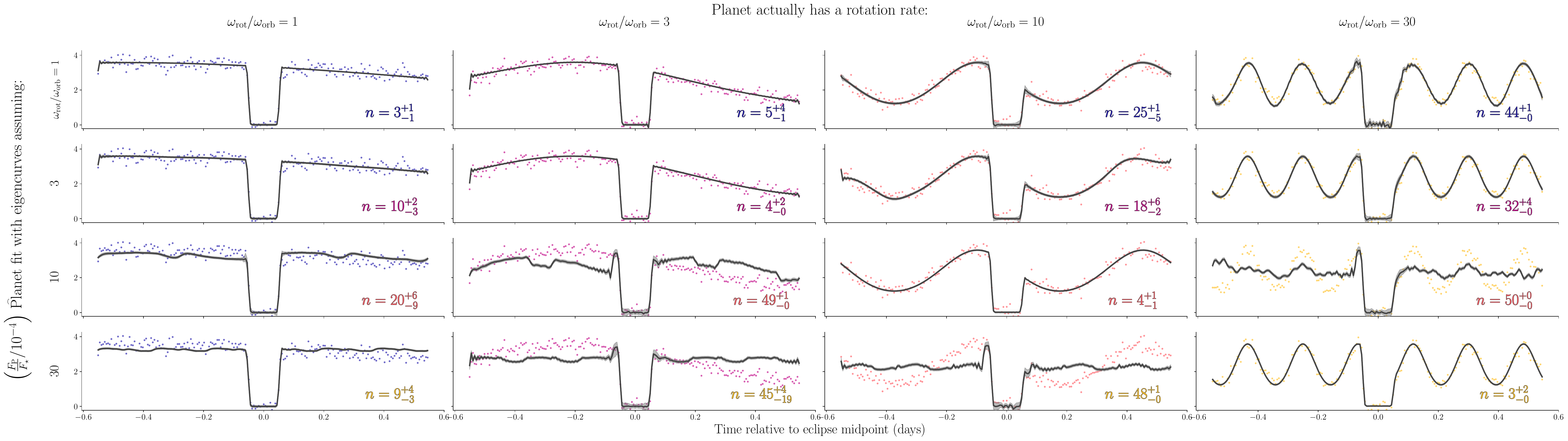} \\
\includegraphics[width=12cm]{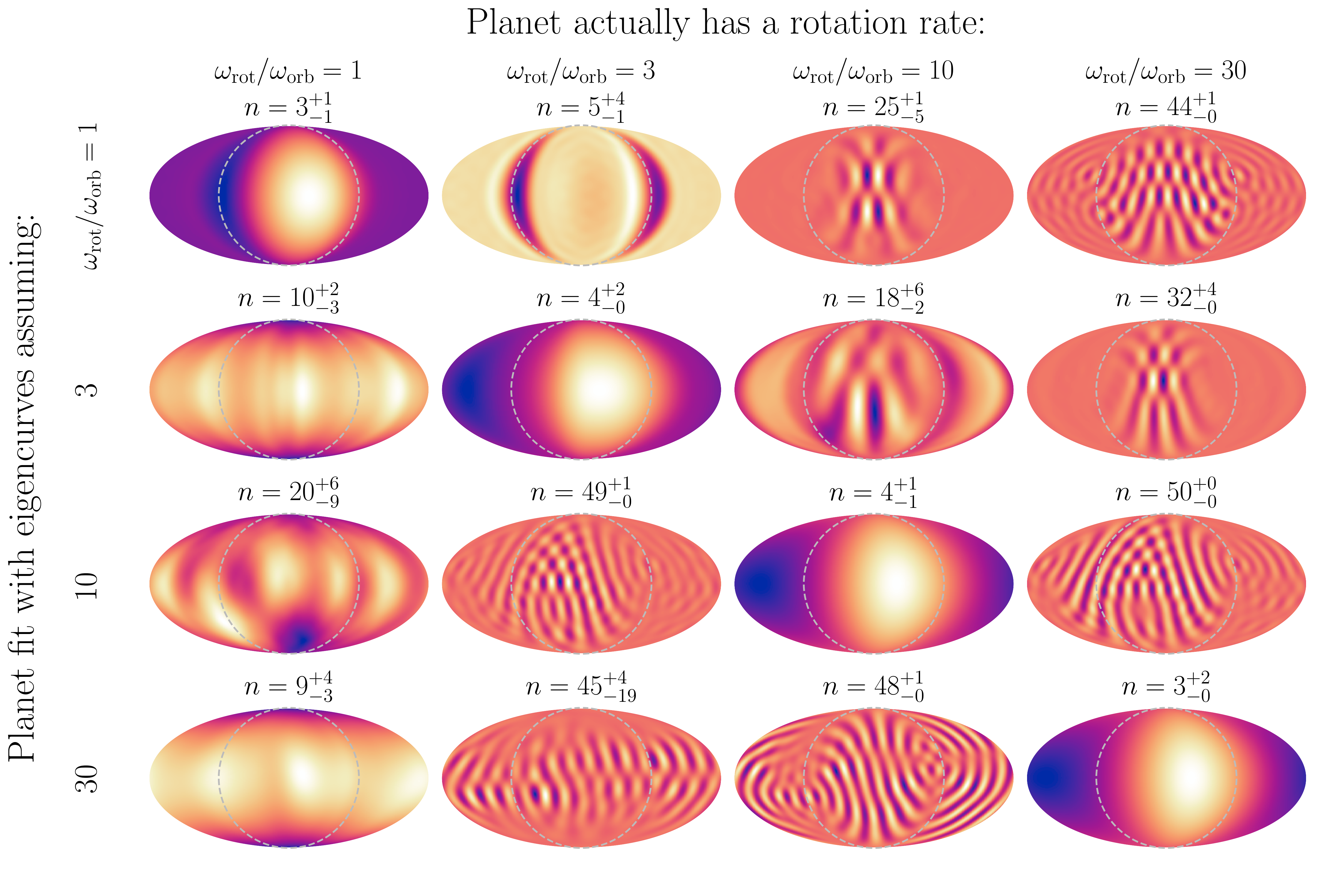} \\
\end{tabular}
\caption{Maps and curves from fits to simulated data from model planets rotating at a range of rates $\omega_\mathrm{rot}/\omega_\mathrm{orb} = 1$--30, when observations span $\pm 10$\% of the orbit centered on secondary eclipse. The input map is shown at the top in green, as it would be oriented to the observer at the time of mid-ingress. The equivalent full-globe Mollweide projection is shown just below to compare with the equivalent projections of the retrieved maps. $\omega_\mathrm{rot}/\omega_\mathrm{orb}$ is equivalent to how many times the planet rotates per orbit. We generate 10000 realizations of Gaussian noise for each simulated phase curve, and fit each with some number $n$ of eigencurves (i.e.~parameters), where $n$ is determined by a Bayesian Information Criterion (BIC). The mode and $\pm 1 \sigma$ of $n$ are shown for each case. Recovered curves and maps are shown based on the rotation rate of the model used to generate the data ($x$-axis), versus the rotation rate of the eigenbasis used to fit the data ($y$-axis). Each sub-plot of the phase curve grid shows the range of uncertainty used to artificially noise the simulated data, with one example realization of the 10000 in colored points. The fit is shown in greyscale; the solid line is the median of all fits and a shaded region shows the $\pm 1 \sigma$ ($\approx$16th and 84th percentiles) range of fits. We show maps generated using the median values of the coefficients from each eigenmap.}
\label{fig:retrieved_zero_obliquity_partial-phase}
\end{center}
\end{figure*}

As a simple example of how this trend can be exaggerated by additional data, we can extend our time series outside of eclipse, to extend to 20\% of the orbital period (still centered around the secondary eclipse, see Figure \ref{fig:retrieved_zero_obliquity_partial-phase}). This would require an additional 25 hours of continuous observation for a single target, a bit expensive for a proposal, but we show it to highlight how, as additional information is gained, we can set a strong prior on the rotation rate for an eclipse mapping fit. Increasing the baseline of observations allows one to separate structure imparted by the eclipsing of map structure, which is only possible during ingress and egress, from structure imparted by rotation, which can be observed across all orbital phases. Note that we have binned to the same number of \emph{bins} here as in the eclipse curve fit, which means we are using the signal from the phase-based rotational variations rather than finely sampled in-eclipse data. As a result, the correlation between map structure and rotation becomes quite strong. In particular, when a spin-synchronous rotation is assumed, the retrieval process creates very artificial brightness structures at high levels of spatial complexity in order to compensate for out-of-eclipse structure. If we have an observed number of waves $n_\mathrm{phase}$ within an ingress or egress, then
\begin{equation}\label{eq:minimum_wavenumber_phase}
    n_\mathrm{phase} \sim \ell_\mathrm{min,lon} \frac{\tau_\mathrm{i,e}}{P_\mathrm{rot}},
\end{equation}
where $\ell_\mathrm{min,lon}$ represents a minimum longitudinal degree of complexity to the map. From the recovered maps in Figure \ref{fig:retrieved_zero_obliquity_partial-phase} we see that at a high enough number of included eigencurves in our fit, we can construct a map that is theoretically ``tuned'' to produce the observed frequency in the phase curve, but at the cost of recruiting higher-degree modes to do so as effectively as is done with just a few modes when the assumed rotation rate is accurate. When the assumed rotation rate of our eigenbasis is much slower than the true rotation rate, the number of eigencurves needed to fit the data push to the extreme of our chosen resolution (here, a maximum of 50 curves). This is an extreme example of where, due to poor constraints in rotation state, the eigenbasis fails in its goal to provide a nearly optimal decomposition of eclipse (and here, phase) curve structure. Here, even if no other prior constraints exist on rotation rate, applying the principle of Occam's razor would provide a strong indication of the scale of rotation that would require the simplest structure to generate.


\subsection{Retrieving Maps with Non-zero Obliquity}\label{sec:retrieval:nonzero-obliquity}
\begin{figure*}[htb!]
\begin{center}
\begin{tabular}{c}
\includegraphics[height=4cm]{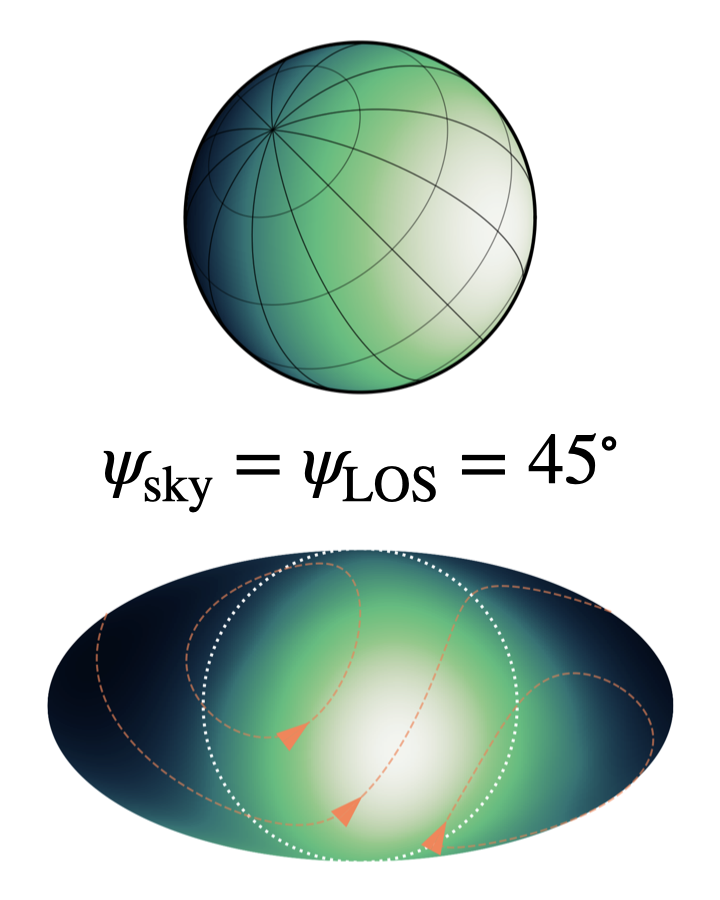} \\
\includegraphics[width=17cm]{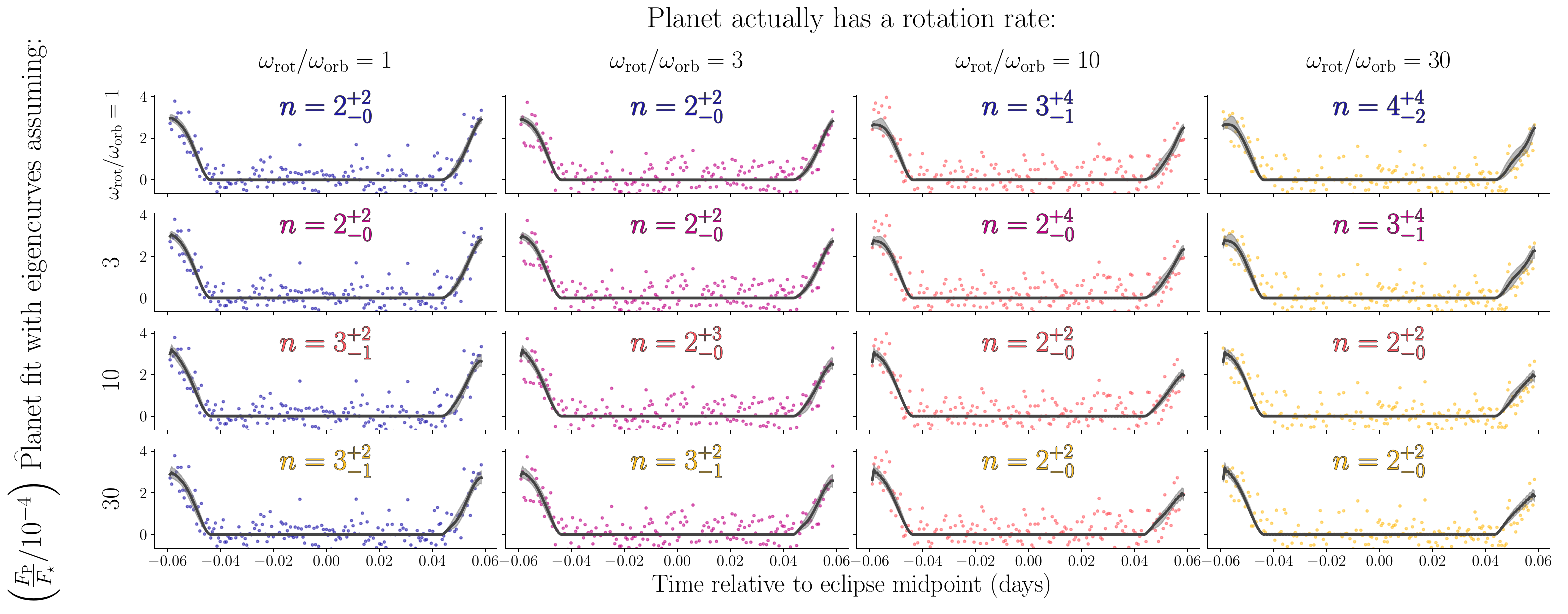} \\
\includegraphics[width=12cm]{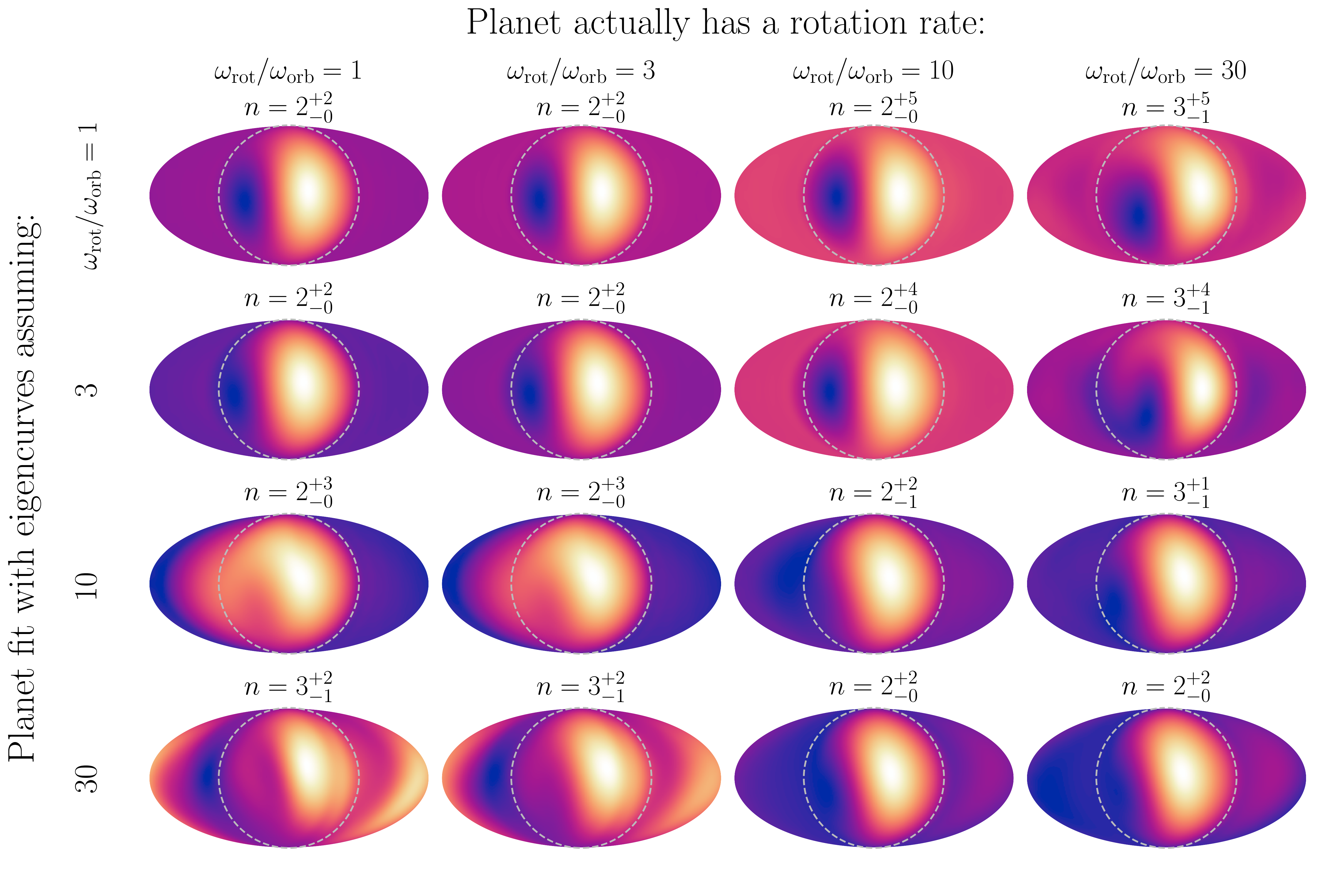} \\
\end{tabular}
\caption{Eclipse maps and curves from fits to simulated data from model planets rotating at a range of rates $\omega_\mathrm{rot}/\omega_\mathrm{orb} = 1$--30, and each with a spin axis that is inclined at 45$^\circ$, both toward the line of sight and with respect to the plane of the sky ($\oblLOS=\oblsky=45^\circ$). The input map is shown at the top in green, as it would be oriented to the observer at the time of mid-ingress. The equivalent full-globe Mollweide projection is shown just below to compare with the equivalent projections of the retrieved maps. $\omega_\mathrm{rot}/\omega_\mathrm{orb}$ is equivalent to how many times the planet rotates per orbit. We generate 10000 realizations of Gaussian noise for each simulated eclipse curve, and fit each with some number $n$ of eigencurves (i.e.~parameters), where $n$ is determined by a Bayesian Information Criterion (BIC). The mode and $\pm 1 \sigma$ of $n$ are shown for each case. Recovered curves and maps are shown based on the rotation rate of the model used to generate the data ($x$-axis), versus the rotation rate of the eigenbasis used to fit the data ($y$-axis). Each sub-plot of the eclipse curve grid shows the range of uncertainty used to artificially noise the simulated data, with one example realization of the 10000 in colored points. The fit is shown in greyscale; the solid line is the median of all fits and a shaded region shows the $\pm 1 \sigma$ ($\approx$16th and 84th percentiles) range of fits. We show maps generated using the median values of the coefficients from each eigenmap.}
\label{fig:retrieved_oblLOS45-oblsky45}
\end{center}
\end{figure*}

In \S \ref{sec:eigenbases} we demonstrated that the structures of our sets of eigencurves look quite similar across a range of obliquities, only diverging as the rotation period becomes comparable with the eclipse durations. As we vary the obliquity of our model HAT-P-18 b planets, and again allow them to rotate at the same range of rates, we see a similar pattern of accuracy in the retrievals. The results are shown in Figure \ref{fig:retrieved_oblLOS45-oblsky45} only for eigenbases that assume zero obliquity; the results are qualitatively similar at this noise level even if we assume the correct obliquity. When the eigenbasis is tuned to the true rotation rate our model is able to converge on the appropriate level of complexity and position for our hotspot map, with just two eigencurves needed. At higher rotations, our ignorance about the rotation rate remains the primary factor in driving up the retrieved complexities of the retrieved maps at higher rotation rates, as opposed to the non-zero obliquities. A minor reduction in retrieved map complexity is however seen at the fastest rotation rate ($\omega_\mathrm{rot}/\omega_\mathrm{orb} = 300$). We see in the upper-right case of our figures (fastest true rotation retrieved with a synchronous eigenbasis) that the distribution of needed eigencurves has a lower median. This is because at higher obliquities, particularly those along the line of sight, the asymmetry in the eclipse data seen with zero obliquity rotation is now more muted. One possible explanation is that, with a sufficiently high value for $\oblLOS$, the amplitude of eclipse curve variations gets muted as now a significant portion of the observed regions on the planet ``persist'' across all rotation phases. That is, there are now points on the planet whose emission is always within the observable disk (``circum-observer'' points). This suggests that obliquity along the line of sight poses the more significant component to contributing to changes in the observed eclipse curves. This also points to why the results do not change qualitatively even when we use an eigenbasis with the correct obliquity angles, in that the information itself is limited, regardless of how well we know the true obliquity, obliquity along the line of sight geometrically limits the amplitude of the signal that is possible to retrieve in this case. An important caveat here is that this is not commenting on the ability of \emph{arbitrary} map structures to be able to produce observable signals --- only that for an equatorial hotspot this type of obliquity ultimately limits our observability. The main conclusion at this level of signal-to-noise is that, at modest to intermediate obliquities, we will be able to retrieve a similar level of map structure as in cases with zero obliquity, and we will be largely insensitive to the precise obliquity unless, as in the cases above, there are independent priors on the rotation state.

\subsection{Retrieving a Latitude-Only Map}\label{sec:retrieval:only-nonzero-obliquity}
\begin{figure*}[htb!]
\begin{center}
\begin{tabular}{c}
\includegraphics[height=4cm]{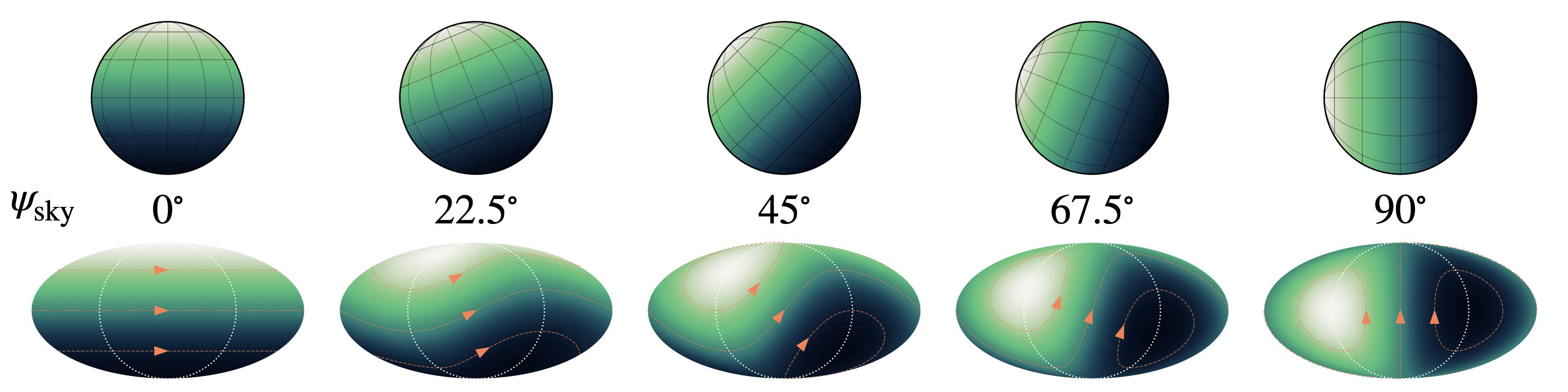} \\
\includegraphics[width=17cm]{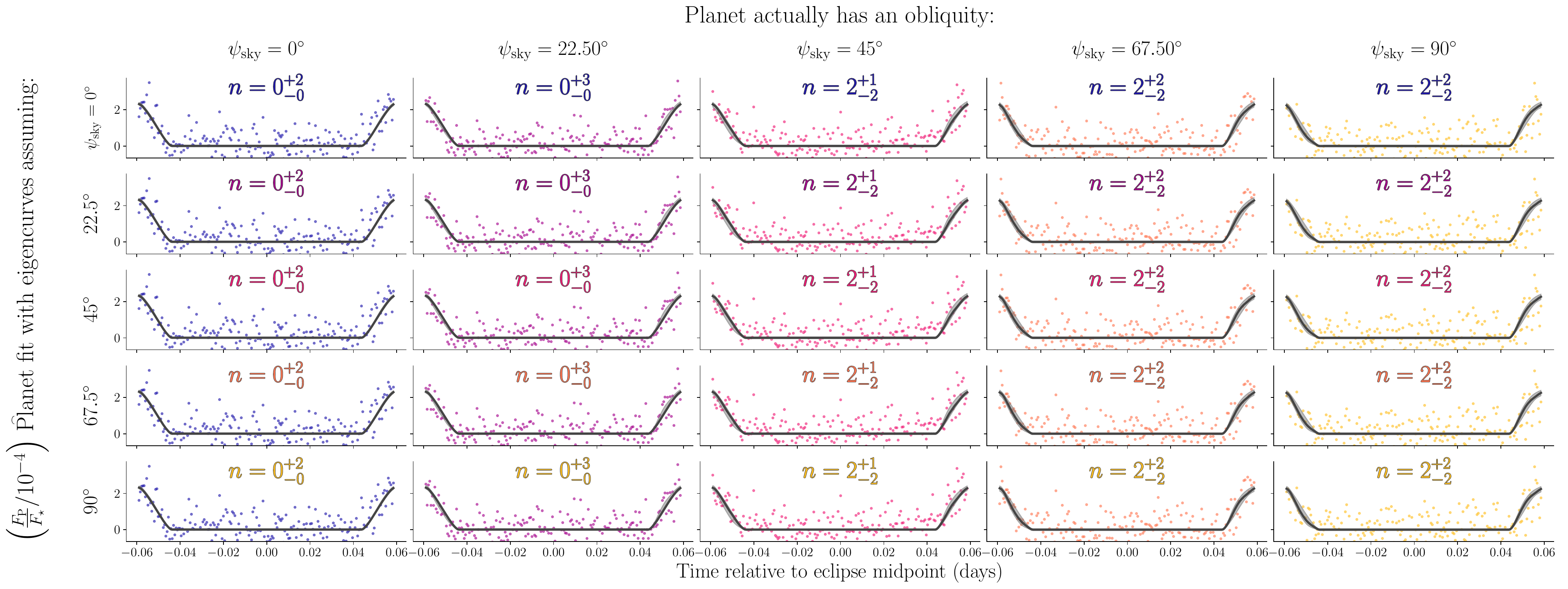} \\
\includegraphics[width=12cm]{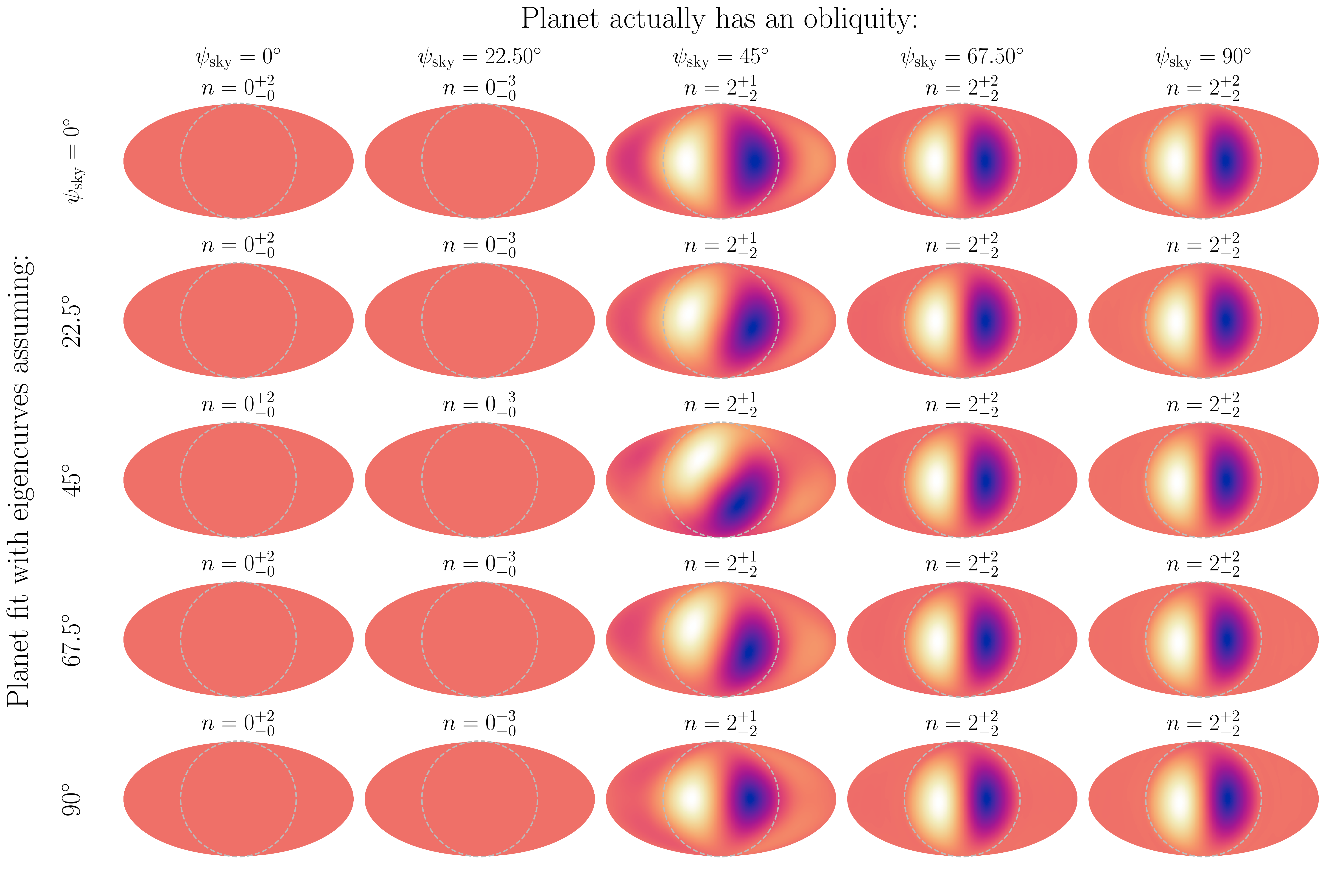} \\
\end{tabular}
\caption{Eclipse maps and curves from fits to simulated data from model planets with polar hotspots with spin axes inclined from 0--90$^\circ$ with respect to the plane of the sky. The input maps are shown at the top in green, as it would be oriented to the observer at the time of mid-ingress. The equivalent full-globe Mollweide projections are shown just below to compare with the equivalent projections of the retrieved maps. $\omega_\mathrm{rot}/\omega_\mathrm{orb}$ is equivalent to how many times the planet rotates per orbit. We generate 10000 realizations of Gaussian noise for each simulated eclipse curve, and fit each with some number $n$ of eigencurves (i.e.~parameters), where $n$ is determined by a Bayesian Information Criterion (BIC). The mode and $\pm 1 \sigma$ of $n$ are shown for each case. Recovered curves and maps are shown based on the rotation rate of the model used to generate the data ($x$-axis), versus the rotation rate of the eigenbasis used to fit the data ($y$-axis). Each sub-plot of the eclipse curve grid shows the range of uncertainty used to artificially noise the simulated data, with one example realization of the 10000 in colored points. The fit is shown in greyscale; the solid line is the median of all fits and a shaded region shows the $\pm 1 \sigma$ ($\approx$16th and 84th percentiles) range of fits. We show maps generated using the median values of the coefficients from each eigenmap.}
\label{fig:retrieved_only-obliquity}
\end{center}
\end{figure*}

We show one final example case building on the analysis in \S \ref{sec:eigenbases:only-obliquity}, by using zero-obliquity eigenbases to retrieve on the simplest non-uniform map that exhibits longitudinal symmetry --- a hemispherical polar hotspot. Here we show how finding the correct obliquity angle can have an effect on the accuracy of retrieved map structure when no information about the rotation rate is available in the eclipse curves. We simulate planets and construct eigenbases across five values of $\oblsky$ from 0--90$^\circ$, as outlined in Table \ref{table:warmJupiter_parameters}. The results of our fitting routine are shown in Figure \ref{fig:retrieved_only-obliquity}. The maximum retrieved map complexity is lower than the previous cases where longitudinal map contrasts could drive eclipse curve variations. However, a clear trend with increasing $\oblsky$ occurs. That is, as the obliquity in the plane of the sky increases, the planet's brightness gradient aligns more favorably with the projected sweep of the occulting edge across the disk of the planet. And, since this alignment is favored during either ingress or egress but not the other, this will produce some reflective asymmetry in the eclipse curve with respect to the midpoint in time. For $\oblsky<45^\circ$, the effect of obliquity is either non-existent or too faint to be retrievable. Once we reach $\oblsky=45^\circ$, this asymmetry is sufficient to allow most of the trials to fit with more than just a uniform map. At this level of signal, the information is mainly limited to detecting the contribution from high sky-plane obliquity, rather than making a constraint. A bit of additional orientation information appears in the case when the true $\oblsky=45^\circ$. Note that, unlike in the basic Warm Jupiter case, the ``ideal'' obliquity value for eclipse structure is roughly $70^\circ$ according to Equation \ref{eq:maximum_gradient}, so while the fact the $\oblsky=45^\circ$ retrieved with an $\oblsky=45^\circ$ may reflect the detection of structure through obliquity, the picture is not conclusive at this noise level. Nevertheless, obliquities along the line of sky can affect how much one can retrieve from planets with longitudinally symmetric map structures.

\section{Conclusions}\label{sec:conclusions}
The eigenbasis method is a powerful tool for optimizing the fitting of eclipse data, providing a structure to analyze the information content across a variety of orbital and rotation configurations. It relies on using principal component analysis to adapt a given basis for spherical maps (spherical harmonics) to a basis which is tuned to the specific observational data available for planets in secondary eclipse. The result is a set of eclipse light curves, a.k.a.~eigencurves, that linearize the fitting and organize the possible map structures that would in theory yield the most observable structures in eclipse, and makes no assumptions or explicit parametrizations about the structures of the maps. By constructing these linearized sets of eclipse eigencurves at various rotation rates and obliquities, we can compare how the most observable structures change with rotation. We then apply these bases to a mock map retrieval using an example JWST observing program of the HAT-P-18 system, where the rotation rate may not necessarily be synchronous. Overall we find that:

\begin{itemize}
    \item Within the broadest range of possible rotation rates and obliquities for Warm Jupiters, and given some possible structure in an eclipse light curve, there is a continuous brightness map that can yield that eclipse curve. That is, the degeneracies one may face in eclipse mapping persist across rotation states --- only the precise projection function changes. However, the specific shapes of these map components may be supported or ruled out with sufficient priors on plausible physical maps, as well as on the rotation.
    \item For most near-term eclipse mapping, where first-order brightness structure is the achievable goal, it is unlikely that even no constraint on rotation rate or obliquity will be the primary source of inaccuracies in retrieving maps. Only if the planet is rotating near its fastest physical limit is there a possibility that a simple structure, e.g.~a hemispherical dipole, would be interpreted as a more complicated map if slow/synchronous rotation is assumed. In this case, exploring a few rotation rates in half-order steps gives an ``Occam's razor'' constraint on rotation rate --- an eigenbasis generated at a faster rotation rate uses fewer eigencurves to fit the data.
    \item We perform the same analysis for planet maps that vary only in latitude, as a simple limiting case for planets where fast rotation may remove any longitudinal structure. In this case the geometric effect of obliquity in the plane of the sky determines the possible signal retrievable through eclipse mapping.
    
\end{itemize}
Our conclusions here are limited to observations of a planet across a single secondary eclipse in a single (arbitrary) photometric band. Additional information such as a partial phase curve can allow for a constraint on the rotation rate of the planet. This is not a true physical constraint but rather an example of how considering the information that is available from mapping can put a strong prior on a physical property; in this case, the rotation rate chosen for the eigenbasis can strongly influence how efficiently the eigenbasis can project maps onto observed curves. This efficiency points to a broader potential exploration of information theory as it applies to exoplanet mapping, as it can help answer a number of questions relevant both in the near- and longer-term. For example, what are the most optimal observational setups beyond a single band in eclipse, as a function of orbital and rotation states? This could include maps which vary in time, such as with large-scale weather patterns similar to Jupiter's Great Red Spot, or maps across multiple wavelengths such as in thermal or albedo mapping. Studies of the information content of these sorts of mapping observations can help hone in on which priors are most impactful. This goes hand in hand with physical models of exoplanet atmospheres which can provide priors on plausible map structures. As one example, \citet{Rauscher2017} used climate modeling to demonstrate that high ($\gtrsim 30^\circ$) obliquities for realistic emission patterns should be detectable with JWST in eclipse mapping. We hope the current work provides a helpful foundation for better understanding the information content of planets as we move beyond mapping Hot Jupiters, which will be increasingly relevant as we are now able to observe planets in and out of eclipse with JWST.

\acknowledgments

We would like to thank Dr. Rodrigo Luger, who helped us understand aspects of the STARRY code in greater detail. We also would like to thank Dr. Bill Blair, who led the development of the JWST example science program referenced in this work. This work was partially funded through an internal MCubed grant from the University of Michigan.

\vspace{5mm}

\software{Astropy \citep{ast13}, Colorcet \citep{kov15}, python-colormath \citep{Taylor2018}, Jupyter \citep{klu16}, Matplotlib \citep{hun07}, Numpy \citep{van11}, Scikit-learn \citep{scikit-learn}, Scipy \citep{jon01}, STARRY \citep{lug19}}

\appendix

\section{Eigenbases at Intermediate Obliquities}\label{sec:appendix}
\begin{figure*}[htb!]
\begin{center}
\includegraphics[width=17cm]{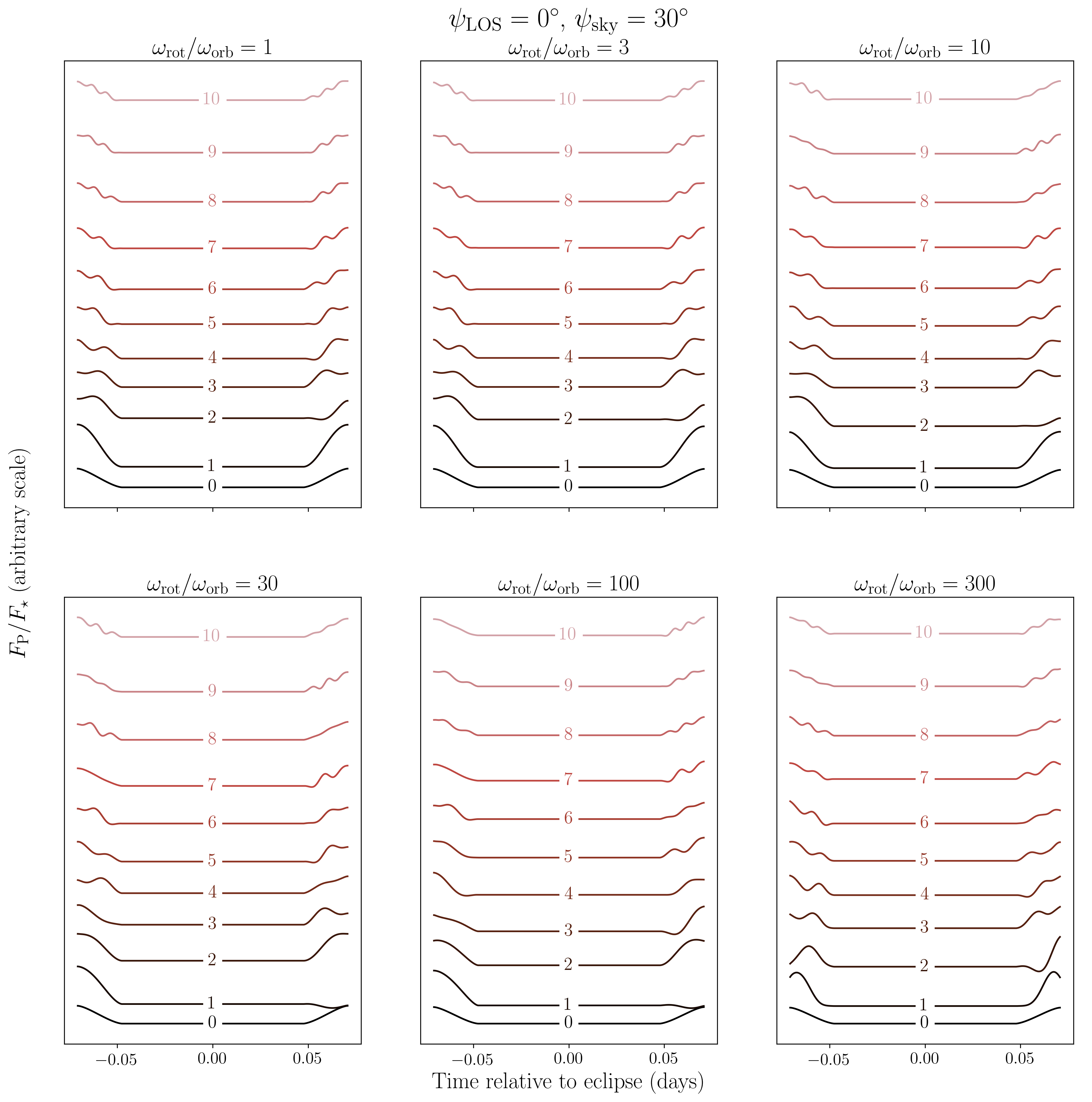}
\caption{The uniform brightness curve and the first 10 eigencurves with uniform component added for our Warm Jupiter model, at a range of rotation rates with $\oblsky = 30^\circ$. Rotation rates span the scale from ``slow'' , where the planet rotates a small fraction of a full rotation across eclipse ($N_\mathrm{tot} < 1$), to ``fast'' where the planet rotates through a considerable fraction of a full rotation. At the fastest rate the planet's rotation period approaches the duration of ingress or egress ($N_\mathrm{i, e} \rightarrow 1$). For reference, with our prototypical 10-day warm Jupiter $N_\mathrm{tot} = 1$ corresponds to $\omega_\mathrm{rot}/\omega_\mathrm{orb} \approx 71$, and $N_\mathrm{i, e} = 1$ corresponds to $\omega_\mathrm{rot}/\omega_\mathrm{orb} \approx 432$. One might notice that some eigencurves dip below the baseline where the planet is completely behind the star and contributes nothing to the observed flux. The eigencurves are initially generated from spherical harmonics, which each have zero net global brightness. Therefore, the eigencurves themselves start out as combinations of maps which each have zero net brightness. To ``reconsitute'' the eigencurves, we plot them here as additions on top of the uniform brightness curve. This allows for an eclipse curve composed of a single eigenmode to be unphysical, i.e.~ have net negative flux at some points.}
\label{fig:obl-sky30_eigencurves}
\end{center}
\end{figure*}

\begin{figure*}[htb!]
\begin{center}
\includegraphics[width=17cm]{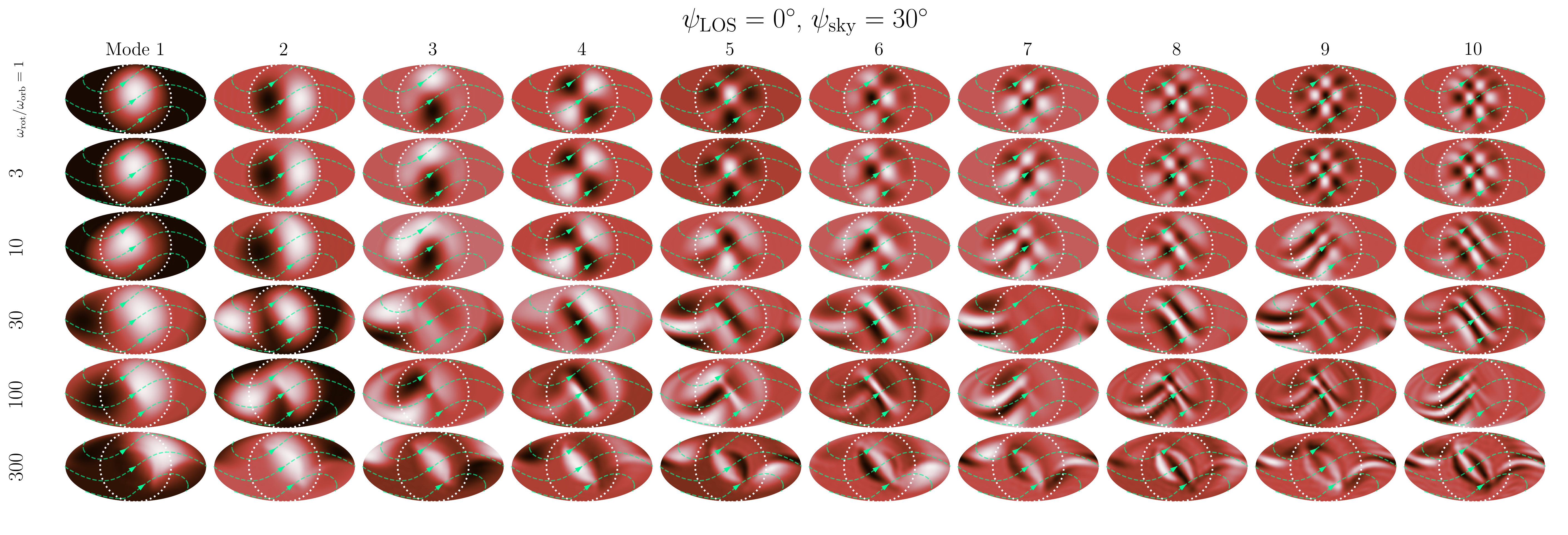}
\caption{The first ten modes of eigenmaps identified by PCA for the Warm Jupiter case with $\oblsky = 30^\circ$. We show the maps in a Mollweide projection that shows all longitudes and latitudes. The reference hemisphere, which is the hemisphere that faces the observer at the mid-point phase of ingress, is centered in each projection and outlined with a dotted white circle. Green dashed contours represent the equator and $\pm 45^\circ$ latitude lines, with arrows indicating the direction of rotation. Rotation rates span the scale from ``slow'' , where the planet rotates a small fraction of a full rotation across eclipse ($N_\mathrm{tot} < 1$), to ``fast'' where the planet rotates through a considerable fraction of a full rotation. At the fastest rate the planet's rotation period approaches the duration of ingress or egress ($N_\mathrm{i, e} \rightarrow 1$). For reference, with our prototypical 10-day warm Jupiter $N_\mathrm{tot} = 1$ corresponds to $\omega_\mathrm{rot}/\omega_\mathrm{orb} \approx 71$, and $N_\mathrm{i, e} = 1$ corresponds to $\omega_\mathrm{rot}/\omega_\mathrm{orb} \approx 432$.}
\label{fig:obl-sky30_eigenmaps}
\end{center}
\end{figure*}

\begin{figure*}[htb!]
\begin{center}
\includegraphics[width=17cm]{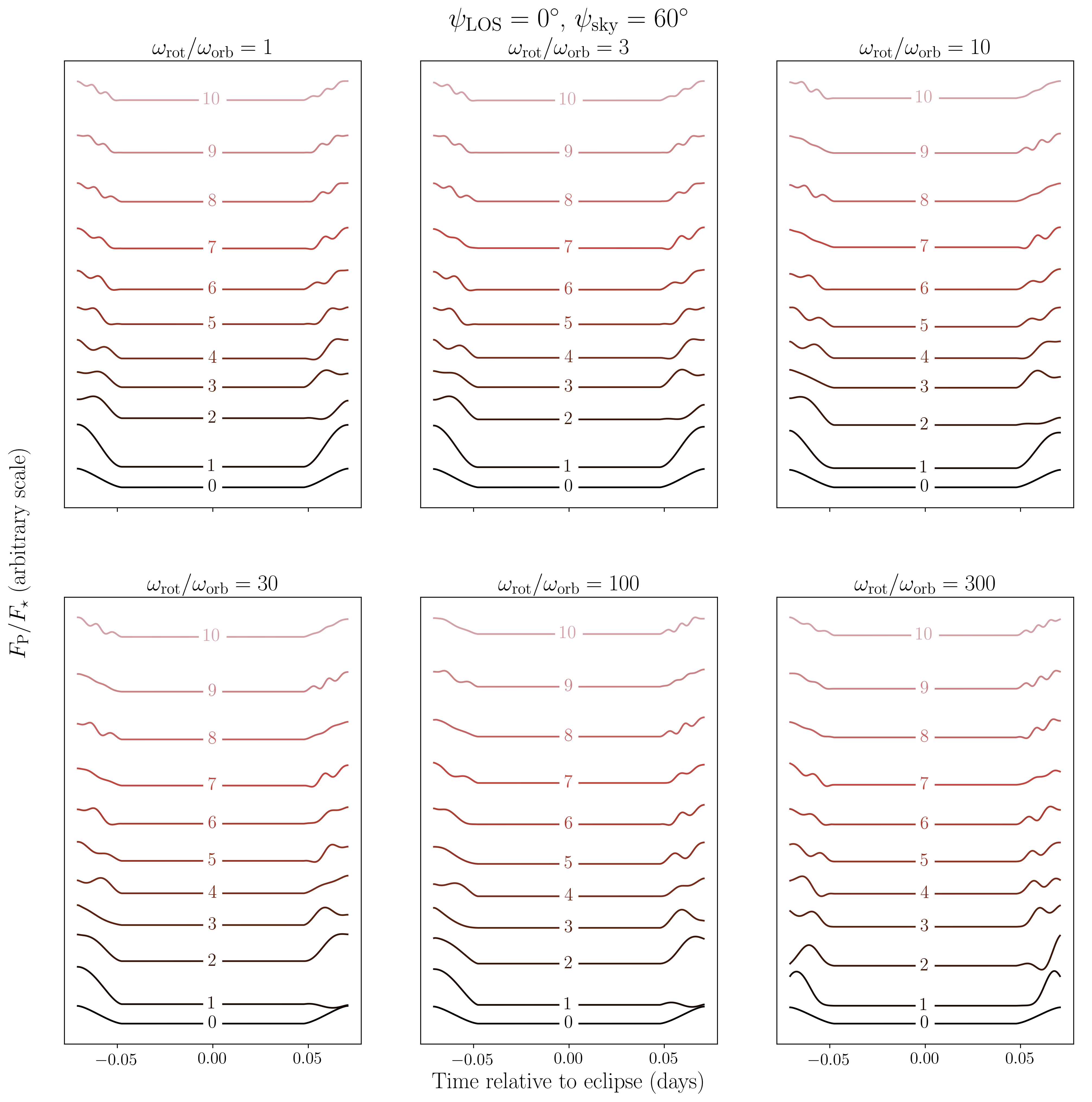}
\caption{The uniform brightness curve and the first 10 eigencurves with uniform component added for our Warm Jupiter model, at a range of rotation rates with $\oblsky = 60^\circ$. Rotation rates span the scale from ``slow'' , where the planet rotates a small fraction of a full rotation across eclipse ($N_\mathrm{tot} < 1$), to ``fast'' where the planet rotates through a considerable fraction of a full rotation. At the fastest rate the planet's rotation period approaches the duration of ingress or egress ($N_\mathrm{i, e} \rightarrow 1$). For reference, with our prototypical 10-day warm Jupiter $N_\mathrm{tot} = 1$ corresponds to $\omega_\mathrm{rot}/\omega_\mathrm{orb} \approx 71$, and $N_\mathrm{i, e} = 1$ corresponds to $\omega_\mathrm{rot}/\omega_\mathrm{orb} \approx 432$. One might notice that some eigencurves dip below the baseline where the planet is completely behind the star and contributes nothing to the observed flux. The eigencurves are initially generated from spherical harmonics, which each have zero net global brightness. Therefore, the eigencurves themselves start out as combinations of maps which each have zero net brightness. To ``reconsitute'' the eigencurves, we plot them here as additions on top of the uniform brightness curve. This allows for an eclipse curve composed of a single eigenmode to be unphysical, i.e.~ have net negative flux at some points.}
\label{fig:obl-sky60_eigencurves}
\end{center}
\end{figure*}

\begin{figure*}[htb!]
\begin{center}
\includegraphics[width=17cm]{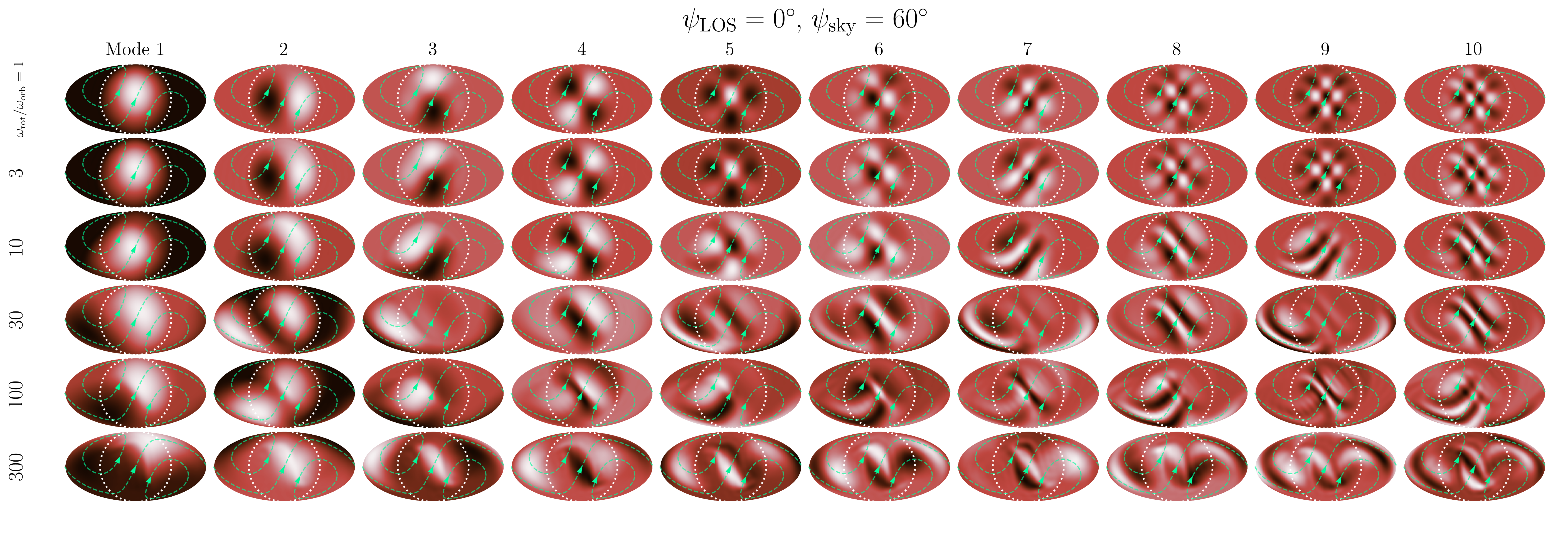}
\caption{The first ten modes of eigenmaps identified by PCA for the Warm Jupiter case with $\oblsky = 60^\circ$. We show the maps in a Mollweide projection that shows all longitudes and latitudes. The reference hemisphere, which is the hemisphere that faces the observer at the mid-point phase of ingress, is centered in each projection and outlined with a dotted white circle. Green dashed contours represent the equator and $\pm 45^\circ$ latitude lines, with arrows indicating the direction of rotation. Rotation rates span the scale from ``slow'' , where the planet rotates a small fraction of a full rotation across eclipse ($N_\mathrm{tot} < 1$), to ``fast'' where the planet rotates through a considerable fraction of a full rotation. At the fastest rate the planet's rotation period approaches the duration of ingress or egress ($N_\mathrm{i, e} \rightarrow 1$). For reference, with our prototypical 10-day warm Jupiter $N_\mathrm{tot} = 1$ corresponds to $\omega_\mathrm{rot}/\omega_\mathrm{orb} \approx 71$, and $N_\mathrm{i, e} = 1$ corresponds to $\omega_\mathrm{rot}/\omega_\mathrm{orb} \approx 432$.}
\label{fig:obl-sky60_eigenmaps}
\end{center}
\end{figure*}

\begin{figure*}[htb!]
\begin{center}
\includegraphics[width=17cm]{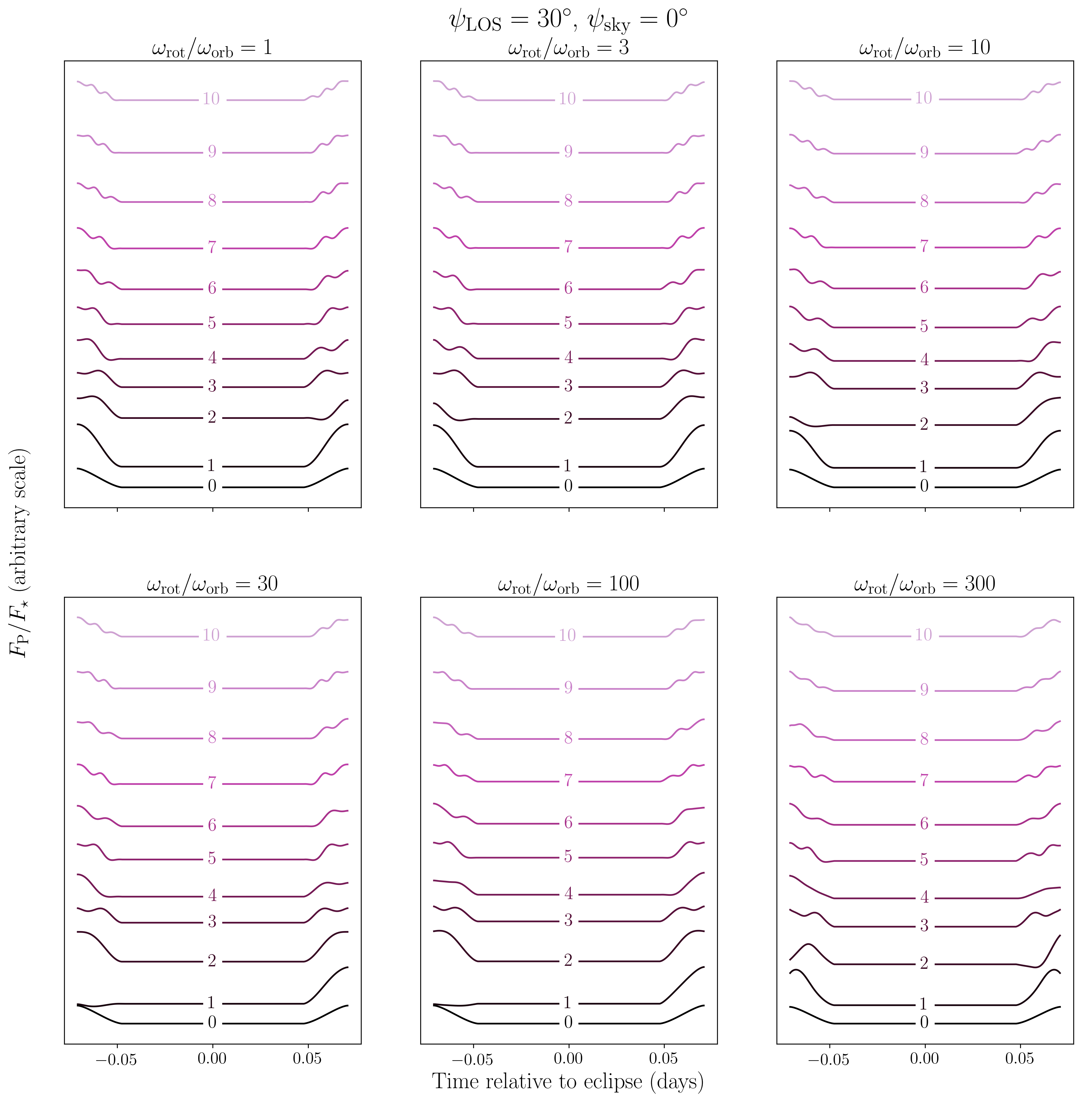}
\caption{The uniform brightness curve and the first 10 eigencurves with uniform component added for our Warm Jupiter model, at a range of rotation rates with $\oblLOS = 30^\circ$. Rotation rates span the scale from ``slow'' , where the planet rotates a small fraction of a full rotation across eclipse ($N_\mathrm{tot} < 1$), to ``fast'' where the planet rotates through a considerable fraction of a full rotation. At the fastest rate the planet's rotation period approaches the duration of ingress or egress ($N_\mathrm{i, e} \rightarrow 1$). For reference, with our prototypical 10-day warm Jupiter $N_\mathrm{tot} = 1$ corresponds to $\omega_\mathrm{rot}/\omega_\mathrm{orb} \approx 71$, and $N_\mathrm{i, e} = 1$ corresponds to $\omega_\mathrm{rot}/\omega_\mathrm{orb} \approx 432$. One might notice that some eigencurves dip below the baseline where the planet is completely behind the star and contributes nothing to the observed flux. The eigencurves are initially generated from spherical harmonics, which each have zero net global brightness. Therefore, the eigencurves themselves start out as combinations of maps which each have zero net brightness. To ``reconsitute'' the eigencurves, we plot them here as additions on top of the uniform brightness curve. This allows for an eclipse curve composed of a single eigenmode to be unphysical, i.e.~ have net negative flux at some points.}
\label{fig:obl-LOS30_eigencurves}
\end{center}
\end{figure*}

\begin{figure*}[htb!]
\begin{center}
\includegraphics[width=17cm]{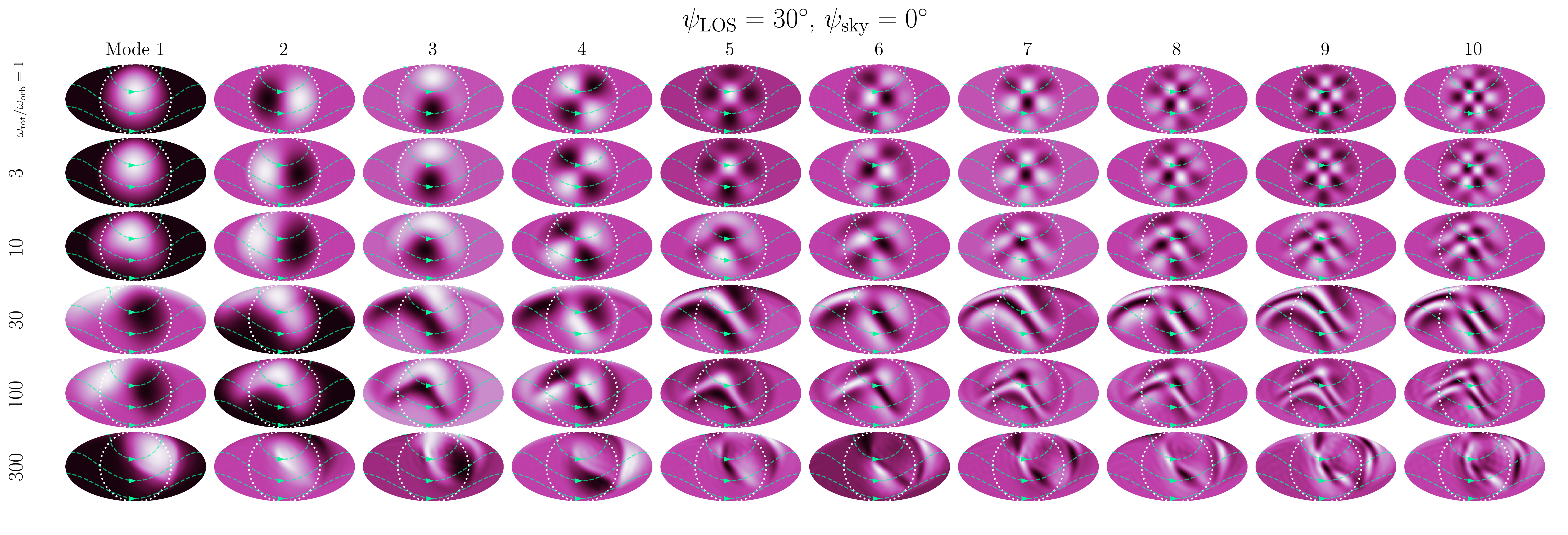}
\caption{The first ten modes of eigenmaps identified by PCA for the Warm Jupiter case with $\oblLOS = 30^\circ$. We show the maps in a Mollweide projection that shows all longitudes and latitudes. The reference hemisphere, which is the hemisphere that faces the observer at the mid-point phase of ingress, is centered in each projection and outlined with a dotted white circle. Green dashed contours represent the equator and $\pm 45^\circ$ latitude lines, with arrows indicating the direction of rotation. Rotation rates span the scale from ``slow'' , where the planet rotates a small fraction of a full rotation across eclipse ($N_\mathrm{tot} < 1$), to ``fast'' where the planet rotates through a considerable fraction of a full rotation. At the fastest rate the planet's rotation period approaches the duration of ingress or egress ($N_\mathrm{i, e} \rightarrow 1$). For reference, with our prototypical 10-day warm Jupiter $N_\mathrm{tot} = 1$ corresponds to $\omega_\mathrm{rot}/\omega_\mathrm{orb} \approx 71$, and $N_\mathrm{i, e} = 1$ corresponds to $\omega_\mathrm{rot}/\omega_\mathrm{orb} \approx 432$.}
\label{fig:obl-LOS30_eigenmaps}
\end{center}
\end{figure*}

\begin{figure*}[htb!]
\begin{center}
\includegraphics[width=17cm]{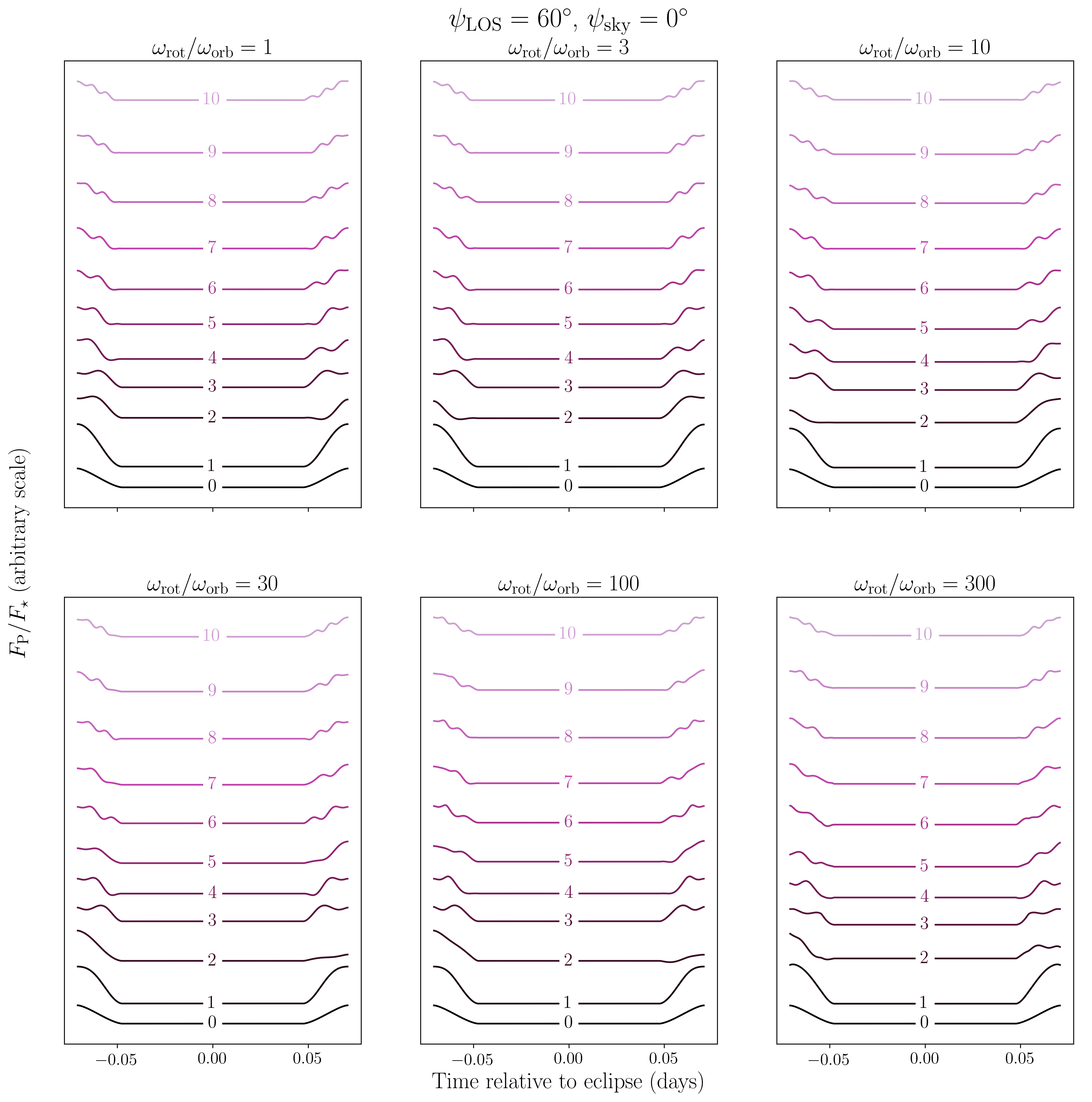}
\caption{The uniform brightness curve and the first 10 eigencurves with uniform component added for our Warm Jupiter model, at a range of rotation rates with $\oblLOS = 60^\circ$. Rotation rates span the scale from ``slow'' , where the planet rotates a small fraction of a full rotation across eclipse ($N_\mathrm{tot} < 1$), to ``fast'' where the planet rotates through a considerable fraction of a full rotation. At the fastest rate the planet's rotation period approaches the duration of ingress or egress ($N_\mathrm{i, e} \rightarrow 1$). For reference, with our prototypical 10-day warm Jupiter $N_\mathrm{tot} = 1$ corresponds to $\omega_\mathrm{rot}/\omega_\mathrm{orb} \approx 71$, and $N_\mathrm{i, e} = 1$ corresponds to $\omega_\mathrm{rot}/\omega_\mathrm{orb} \approx 432$. One might notice that some eigencurves dip below the baseline where the planet is completely behind the star and contributes nothing to the observed flux. The eigencurves are initially generated from spherical harmonics, which each have zero net global brightness. Therefore, the eigencurves themselves start out as combinations of maps which each have zero net brightness. To ``reconsitute'' the eigencurves, we plot them here as additions on top of the uniform brightness curve. This allows for an eclipse curve composed of a single eigenmode to be unphysical, i.e.~ have net negative flux at some points.}
\label{fig:obl-LOS60_eigencurves}
\end{center}
\end{figure*}

\begin{figure*}[htb!]
\begin{center}
\includegraphics[width=17cm]{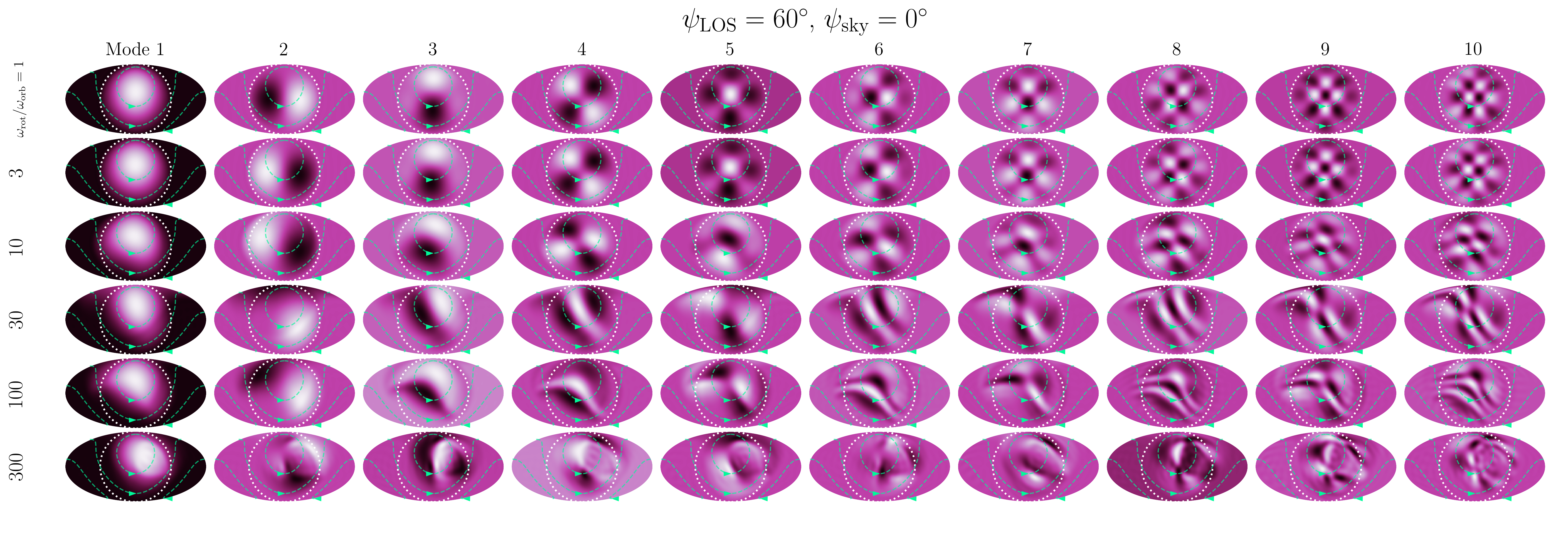}
\caption{The first ten modes of eigenmaps identified by PCA for the Warm Jupiter case with $\oblLOS = 60^\circ$. We show the maps in a Mollweide projection that shows all longitudes and latitudes. The reference hemisphere, which is the hemisphere that faces the observer at the mid-point phase of ingress, is centered in each projection and outlined with a dotted white circle. Green dashed contours represent the equator and $\pm 45^\circ$ latitude lines, with arrows indicating the direction of rotation. Rotation rates span the scale from ``slow'' , where the planet rotates a small fraction of a full rotation across eclipse ($N_\mathrm{tot} < 1$), to ``fast'' where the planet rotates through a considerable fraction of a full rotation. At the fastest rate the planet's rotation period approaches the duration of ingress or egress ($N_\mathrm{i, e} \rightarrow 1$). For reference, with our prototypical 10-day warm Jupiter $N_\mathrm{tot} = 1$ corresponds to $\omega_\mathrm{rot}/\omega_\mathrm{orb} \approx 71$, and $N_\mathrm{i, e} = 1$ corresponds to $\omega_\mathrm{rot}/\omega_\mathrm{orb} \approx 432$.}
\label{fig:obl-LOS60_eigenmaps}
\end{center}
\end{figure*}

For a more complete picture of the transition between the bases of planet maps when obliquity is increased, we provide the eigencurves and eigenmaps for the intermediate obliquities ($30^\circ$ and $60^\circ$) for both $\oblsky$ (Figures \ref{fig:obl-sky30_eigencurves}--\ref{fig:obl-sky60_eigenmaps}) and $\oblLOS$ (Figures \ref{fig:obl-LOS30_eigencurves}--\ref{fig:obl-LOS60_eigenmaps}).

\bibliography{library}{}
\bibliographystyle{aasjournal}

\end{document}